\documentclass[a4paper,fleqn,usenatbib]{mnras}
\usepackage{newtxtext,newtxmath}
\usepackage{ae,aecompl}
\usepackage{graphicx}	
\usepackage{amsmath}	
\usepackage{amssymb}	
\usepackage{mathtools}
\usepackage[applemac]{inputenc}
\usepackage[T1]{fontenc}
\usepackage{lmodern}
\usepackage{textcomp}
\usepackage{hyperref}
\usepackage{natbib}
\usepackage[english]{babel}
\usepackage{fixltx2e}
\usepackage{subfigure}
\usepackage[labelfont=bf]{caption}
\usepackage{lineno}
\usepackage{times}

\usepackage{fancyhdr}

\usepackage{color}

\def\be{\begin{equation}}
\def\ee{\end{equation}}
\def\ba{\begin{array}{lll}}
\def\ea{\end{array}}
\def\beq{\begin{eqnarray}}
\def\eeq{\end{eqnarray}}

\title[SNRs in clumpy media]{Supernova remnants in clumpy media: particle propagation and gamma-ray emission}

\author[]{
S.~Celli, $^{1}$  \thanks{silvia.celli@gssi.infn.it}
G.~Morlino,$^{1,2}$ 
S.~Gabici$^{3}$ 
and F.~A.~Aharonian$^{1,4,5}$ 
\\
$^{1}$Gran Sasso Science Institute, Viale Francesco Crispi 7, 67100, L'Aquila, Italy \\
$^{2}$INAF, Osservatorio Astrofisico di Arcetri, L.go E.~Fermi 5, Firenze, Italy  \\
$^{3}$APC, AstroParticule et Cosmologie, Universit\'e Paris Diderot, CNRS/IN2P3, CEA/Irfu, Obs de Paris, Sorbonne Paris Cit\'e, France \\
$^{4}$Max-Planck-Institut f\"ur Kernphysik, Postfach 103980, D-69029 Heidelberg, Germany\\
$^{5}$Dublin Institute for Advanced Studies, 31 Fitzwilliam Place, Dublin 2, Ireland  \\
}

\date{Accepted XXX. Received YYY; in original form ZZZ}

\pubyear{2019}

\begin{document}
\label{firstpage}
\pagerange{\pageref{firstpage}--\pageref{lastpage}}
\maketitle

\begin{abstract}
Observations from the radio to the gamma-ray wavelengths indicate that supernova remnant (SNR) shocks are sites of effective particle acceleration.
It has been proposed that the presence of dense clumps in the environment where supernovae explode might have a strong impact in shaping the hadronic gamma-ray spectrum.
Here we present a detailed numerical study about the penetration of relativistic protons into clumps which are engulfed by a SNR shock, taking into account the magneto-hydrodynamical properties of the background plasma.
We show that the spectrum of protons inside clumps is much harder than that in the diffuse inter-clump medium and we discuss the implications for the formation of the spectrum of hadronic gamma rays, which does not reflect anymore the acceleration spectrum of protons, resulting substantially modified inside the clumps due to propagation effects.
For the Galactic SNR RX J1713.7-3946, we show that a hadronic scenario including dense clumps inside the remnant shell is able to reproduce the broadband gamma-ray spectrum from GeV to TeV energies.
Moreover, we argue that small clumps crossed by the shock could provide a natural explanation to the non-thermal X-ray variability observed in some hot spots of RX J1713.7-3946.
Finally we discuss the detectability of gamma-ray emission from clumps with the upcoming Cherenkov Telescope Array and the possible detection of the clumps themselves through molecular lines.
\end{abstract}



\section{Introduction}
\label{sec:Intro}

It is now well established that supernova remnants can accelerate electrons and presumably also hadrons, as demonstrated by the non-thermal emission detected from the great majority of SNRs. In particular, the pion-bump detected at $\approx 280$ MeV from two middle-aged SNRs, IC~443 and W~44, indicates the presence of accelerated hadrons \citep{Ackermann2013} (but see also \citet{cardillo} for an alternative explanation). Nevertheless two main questions need to be answered in order to validate the idea that SNRs represent the main source of the Galactic cosmic rays (CRs) observed at Earth: which is the total amount of energy channeled into relativistic particles and which is the final energy spectrum of accelerated particles injected into the interstellar medium (ISM). Gamma-ray observations provide a powerful tool to answer these questions, allowing to directly infer the properties of accelerated hadrons.

In some young SNRs it is still unclear whether the detected gamma-ray emission is produced by leptonic processes, via inverse Compton (IC) scattering, or hadronic collisions, through the decay of the resulting neutral pions. On a general ground the two scenarios mainly differ in the required magnetic field strength: the IC scenario usually requires a very low magnetic field (of the order of $10\,\mu$G) in order to simultaneously account for the radio, X-ray and gamma-ray emission, while the hadronic scenario requires much larger values, of the order of few hundreds of $\mu$G. Such a large magnetic field can not result from the simple compression of the interstellar magnetic field, but requires some amplification, which is, in turn, a possible signature of efficient CR acceleration itself. 
The case of RX J1713.7-3946 is of special interest to this respect. This remnant has been considered for long time the best candidate for an efficient acceleration scenario, mainly due to its high gamma-ray flux. The detection of gamma-ray emission in the energy range $[1-300]$~GeV by the Fermi-LAT satellite (\citealp{fermi}) has shown an unusually hard spectrum which, at a first glance, seems to be in a better agreement with a leptonic scenario. Nevertheless, a deeper analysis shows that neither the hadronic nor the leptonic scenarios, taken in their simplest form, can unequivocally explain the observations \citep{Morlino2009,felix,GabiciAharonian2016}. To address this issue, it has been proposed by \citet{felix}, \citet{inoue} and \citet{stefanofelix} that the very hard energy spectrum at low energies could result also from hadronic emission if the SNR is expanding inside a clumpy medium. In the presence of dense inhomogeneities, in fact, the magnetic field can result amplified all around the clump because of both the field compression and the magneto-hydrodynamical (MHD) instabilities developed in the shock-clump interaction: this makes difficult for particles at low energies to penetrate inside the clump compared to the most energetic ones. Consequently, the gamma-ray spectrum would be harder than the parent proton spectrum accelerated at the forward shock. 

In this work we study the realization of such a scenario. In \S~\ref{sec:clump} we introduce clumps and describe the interaction between a shock and a clump, showing the temporal and spatial evolution of the background plasma properties through a 3D MHD simulation. Then, in \S~\ref{sec:transport}, we focus on the temporal and spatial evolution of the density of accelerated particles, solving the 3D transport equation for CRs propagating into a clumpy medium where a shock is expanding. The diffusion coefficient around the clump is parametrized in order to reproduce the amplified magnetic field resulting from both the regular field compression and MHD instabilities. Once the behavior of the particle density is obtained, the CR spectrum resulting from inside and outside the clump region is derived, showing that the spectrum inside the clump is much harder at early epochs and steepens at later times. The same spectral trend is exhibited in the related gamma-ray emission from inelastic proton-proton collisions, as shown in \S~\ref{sec:gamma}. In order to move from the gamma-ray emission of individual clumps to their cumulative contribution, we will assume a uniform spatial distribution of clumps. The case of RX J1713.7-3946 is then presented in \S~\ref{sec:1713}: in particular, in \S~\ref{subsec:mlines} we explore the possibility to detect clumps embedded inside the remnant through molecular lines, while in \S~\ref{subsec:xray} we comment on the fast variability observed in non-thermal X rays. Finally, in \S~\ref{subsec:CTA}, we discuss the detectability of clump-shock associations by the major next generation ground-based gamma-ray instrument, the Cherenkov Telescope Array (CTA), and we summarize conclusions in \S~\ref{sec:concl}. 

\section{Shock propagation through a clumpy medium}
\label{sec:clump}

Observations of the ISM have revealed a strong non homogeneity, particularly inside the Galactic Plane. On the scales of the order of few parsecs, dense molecular clouds (MCs) constitute structures, mainly composed of H$_2$ molecules, while their ionized component is composed by C ions. Typical temperatures and masses are respectively of the order of $T_\textrm{MC} \simeq 10^2$~K and $M_\textrm{MC} \gtrsim 10^3 \, \textrm{M}_\odot$. On smaller scales (of the order of a fraction of a parsec), colder and denser molecular clumps are present: they are characterized by typical temperatures of $T_\textrm{c} \simeq 10$~K, masses of the order of $M_\textrm{c} \simeq 0.1-1 \, \textrm{M}_\odot$ and therefore number densities of the order of $n_\textrm{c} \gtrsim 10^3$~cm$^{-3}$. In the following, we will consider the lower bound as a reference value for the density of target gas inside individual clumps. These clumps are mostly composed by neutral particles while the most relevant ions are HCO$^+$. Typical values for the ionized density are at least equal to $n_\textrm{i} \leq 10^{-4} n_\textrm{c}$, while the average mass of ions is $m_\textrm{i}=29m_\textrm{p}$ ($m_\textrm{p}$ is the proton mass) and that of neutrals is $m_\textrm{n}=2m_\textrm{p}$. Therefore the ion to neutral mass density $\epsilon$ in clumps amounts to \citep{gabicimontmerle}
\begin{equation}  \label{eq:epsilon}
  \epsilon = \frac{m_\textrm{i} n_\textrm{i}}{m_\textrm{n} n_\textrm{n}} \leq 1.5 \times 10^{-3} \ll 1
\end{equation}
Molecular clouds are strongly influenced by the presence of cosmic rays, since most likely low-energy CRs provide their ionization rate (see \citealp{padovani, Phan2018}), which in turns controls both the chemistry of clouds and the coupling of plasma with local magnetic fields, and hence star formation processes. 

Shocks propagating through inhomogeneities of the ISM are able to generate MHD instabilities, which modify the thermal properties of the plasma and might be able to disrupt the clumps because of thermal conduction \citep{orlando2008}. Therefore the dynamical interaction between the shock emitted at the supernova (SN) explosion and the medium surrounding the star is an essential ingredient for the understanding of star formation processes \citep{hennebelle, dwarkadas}. In particular, the environment where type II SNe explode is most likely populated by molecular clumps: indeed, given their fast evolution, they explode in an environment rich of molecular clouds, the same that generated the star. Moreover, given their massive progenitor, strong winds in the giant phase of the star evolution accelerate the fragmentation of clouds into clumps, while creating a large cavity of hot and rarified gas around them. The physical size of ISM inhomogeneities results from HD simulations which include interstellar cooling, heating and thermal conduction. These have been conducted by \citet{inoue}, who found that the characteristic length scale of clumps amounts to $R_c \simeq 0.1$~pc, corresponding to the smaller scale where thermal instability is effective. We fix this scale throughout the paper. This scenario could be similar to the one in which the remnant of RX J1713.7-3946 is evolving \citep{slane}. Previous works in this direction \citep{klein} have shown that plasma instabilities arise when the upstream medium is not homogeneous \citep{giacalone, fraschetti, sano12}. In particular, in the presence of density inhomogeneities, vorticity may develop after the passage of the shock. In such a situation, the Kelvin-Helmotz instability arises, due to the velocity shear among the clumps and the surrounding medium, generating a strong enhancement of the local magnetic field. The generated turbulence then cascades to smaller spatial scales through a Kolmogorov-like process \citep{inoue} on time scales $\tau_\textrm{cascade}=R_c/v_A \simeq 50$~yr (where the Alfven speed of MHD waves $v_A$ is computed in a low density environment as that of a rarefied cavity). In the next Section, using MHD simulations, a simplified scenario is studied where a single clump much denser than the circumstellar plasma is engulfed by a shock. In such a configuration, two different kinds of processes produce magnetic field amplification. Around the clump, the regular field results to be compressed and stretched reaching a value up to $\sim 10$ times the original value, in a layer that is found to be about half of the clump size, i.e. $0.05$~pc, similar to the results obtained by \citet{inoue}. This region will be referred to as the \emph{clump magnetic skin}. In addition, in the region behind the clump, turbulence develops and the associate vorticity further amplifies the magnetic field.
It is also shown that if the density contrast between the clump and the surrounding medium is very large, the clump can survive for long time before evaporating, even longer than the SNR age.

\subsection{MHD simulations of the shock-clump interaction}
\label{subsec:mhd}
The description of the thermal properties of a classical fluid follows from the solution of the Navier-Stokes equations, coupled to the induction equation for the time evolution of the background magnetic field $\mathbf{~B_0}$. The MHD equations of motion to be solved read, for non resistive fluids, as
\begin{equation}
\label{eq:mhd}
\begin{dcases}
\frac{\partial \rho}{\partial t} + \mathbf{\nabla} \cdot (\rho \mathbf{v})=0  \\
\frac{\partial \mathbf{v}}{\partial t} + \mathbf{v} \cdot \mathbf{\nabla} \mathbf{v} = - \frac{1}{\rho} \left( \mathbf{\nabla} p + \frac{1}{4\pi} (\mathbf{\nabla} \times \mathbf{B_0}) \times \mathbf{B_0} \right) \\
\frac{\partial}{\partial t} \left( \mathcal{E} + \frac{B_0^2}{8\pi} \right) = - \mathbf{\nabla} \cdot \left[ \left( \mathcal{E} + p \right) \mathbf{v} + \frac{1}{4\pi} \mathbf{B_0} \times (\mathbf{v} \times \mathbf{B_0}) \right] \\
\frac{\partial \mathbf{B_0}}{\partial t} = \mathbf{\nabla} \times (\mathbf{v} \times \mathbf{B_0})  \\ 
\mathbf{\nabla} \cdot \mathbf{B_0} = 0
\end{dcases}
\end{equation}
where $\rho$ is the fluid density, $\mathbf{v}$ its velocity, $p$ its pressure, $\mathcal{E}=\rho v^2/2 + p/(\gamma-1)$ its kinetic plus internal energy and $\gamma$ its adiabatic index ($\gamma=5/3$ for an ideal monoatomic non-relativistic gas).\\
In order to introduce a shock discontinuity, as well as the presence of a clump, a numerical approach has been adopted, though the \text{PLUTO} code (see \citealp{pluto}). The shock-cloud interaction is implemented among one of the possible configurations provided by this code. We are interested in performing a three dimensional simulation, in cartesian coordinates. The finite difference scheme adopted for the solution of Eq.~(\ref{eq:mhd}) is based on an unsplit $3^\textrm{rd}$ order Runge-Kutta algorithm with an adaptive time step subject to the Courant condition $C = 0.3$. In order to investigate a situation as much similar as that of high mass star SN explosion, like RX J1713.7-3946, we set in the upstream region a low density medium with n$_\textrm{up} = 10^{-2}$~cm$^{-3}$, which gets compressed by the shock in the downstream region up to n$_\textrm{down} = 4 \times 10^{-2}$~cm$^{-3}$. A strong shock is moving in the direction of the $z$-axis, with a sonic Mach number $M = v_s/c_s \simeq 37$, as expected for this remnant, given its measured shock speed of $v_s = 4.4 \times 10^8$~cm~s$^{-1}$ \citep{uchiyama} and the upstream temperature of $T=10^6$~K, which is typical for bubbles inflated by stellar winds. A flat density profile clump is set in the upstream as initial condition, with a density as high as n$_\textrm{c} = 10^{3}$~cm$^{-3}$. Therefore, a density contrast 
\begin{equation}
\label{eq:chi}
\chi = \frac{n_c}{n_\textrm{up}} = 10^5
\end{equation}
is assumed: if the shock speed is $\mathbf{v}=v_s \widehat{z}$, the shock propagates inside the clump with a velocity $\mathbf{v_{s,c}}=v_{s,c} \widehat{z}$ equal to (\citealp{klein} and \citealp{inoue})
\begin{equation}
\label{eq:vsc}
v_{s,c} = \frac{v_s}{\sqrt{\chi}} = 1.4 \times 10^6 \, \textrm{cm} \, \textrm{s}^{-1}
\end{equation}
Boundary conditions are set as outflow in all directions, except for the downstream boundary in the $z$-direction, where an injection flow is set. We set a uniform grid with size $2$~pc$\times 2$~pc$\times 2$~pc and spatial resolution of $\Delta x =\Delta y = \Delta z = 0.01$~pc. A spherical clump of radius R$_c=0.1$~pc is located in $x_0=y_0=z_0=1$~pc. All the evolution is followed in the clump reference frame. 

In the MHD simulation the clump is assumed to be fully ionized: this is not the real condition, since molecular clumps are mainly composed by neutrals, as discussed in \S~\ref{sec:clump}. However, while the shock is passing through the ionized part of the clump, the heated ions are able to ionize the neutral part on a time scale of the order of few years. This condition allows us to consider the neutral clump as if it were completely ionized. In this process ions cool down and the pressure drops accordingly, hence reducing the shock speed to the a value given by $v_{s,c}$ \citep{klein}. The time needed for the shock to cross the clump is the so called clump crossing time
\begin{equation}
\label{eq:taucc}
\tau_{cc}=\frac{2R_c}{v_{s,c}} \simeq 1.4 \times 10^4 \, \textrm{yr}
\end{equation}
It has been shown through both analytical estimates \citep{klein, chevalier} and simulations \citep{orlando2005, orlando2008} that the time required for the clump to evaporate is of the order of few times $\tau_{cc}$.
This timescale is larger than the estimated age of RX J1713.7-3946, which amounts to $T_\textrm{SNR}\simeq 1625$ yr.

\begin{figure*}
\centering
\includegraphics[scale=0.52]{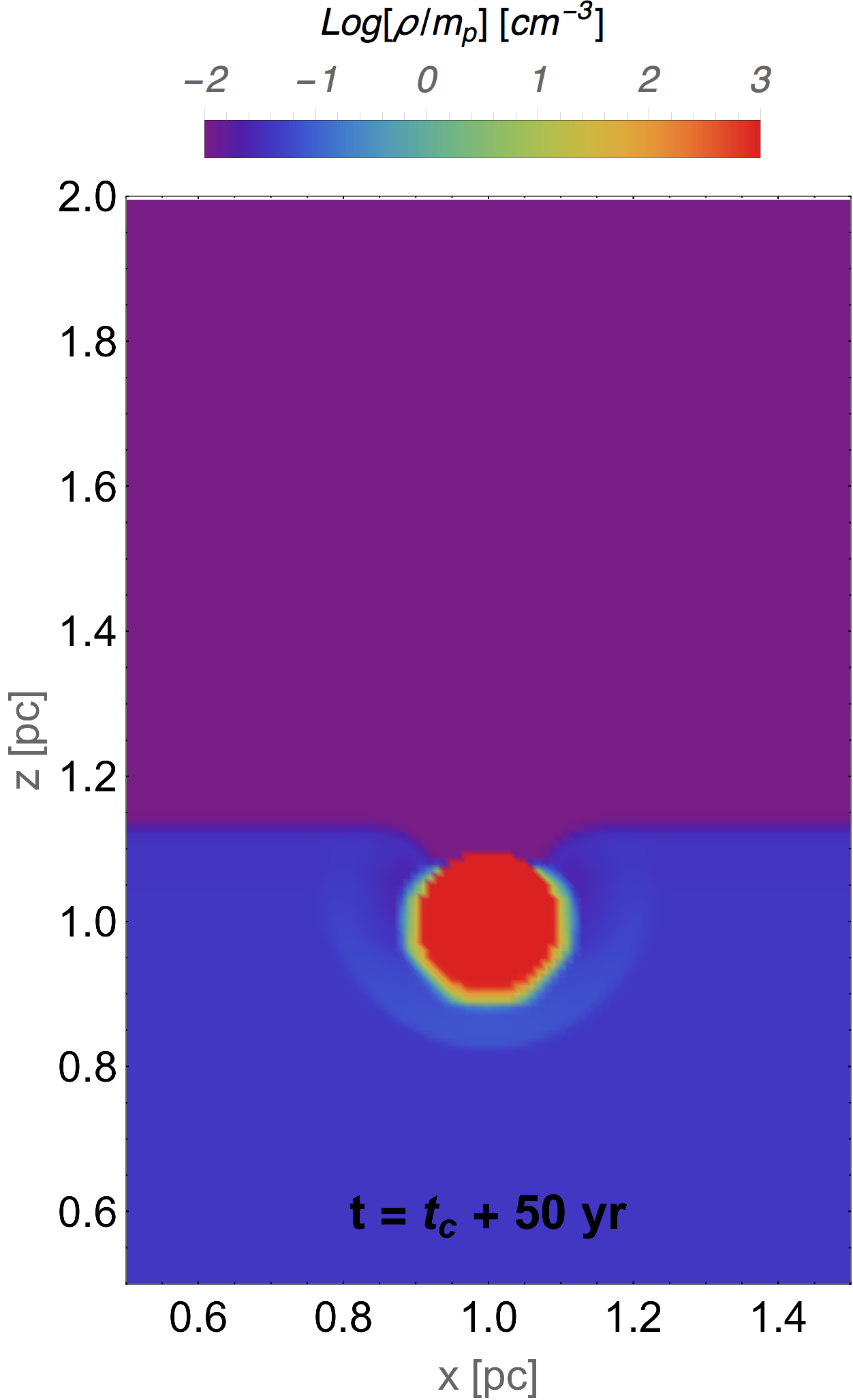}
\includegraphics[scale=0.52]{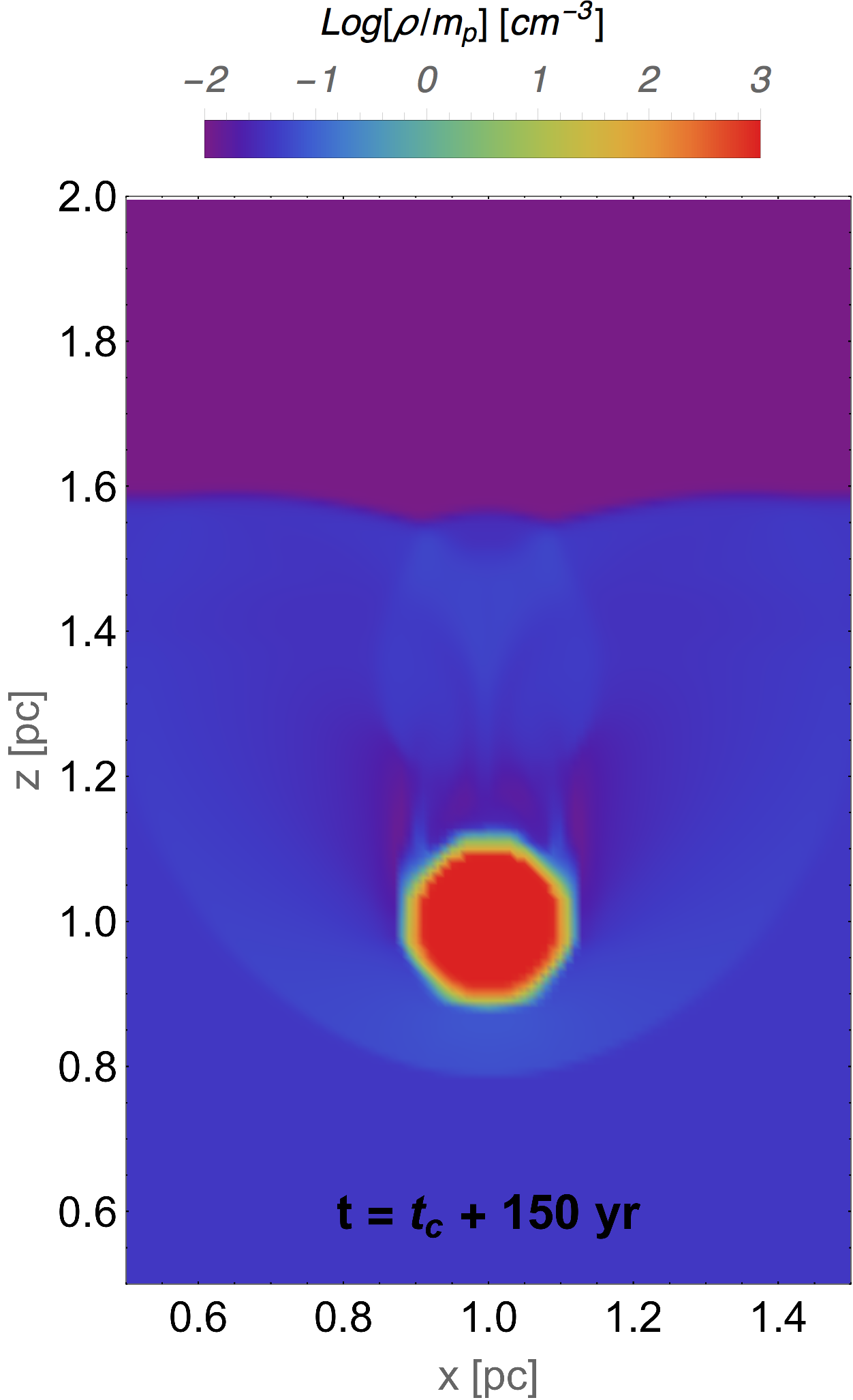}
\includegraphics[scale=0.52]{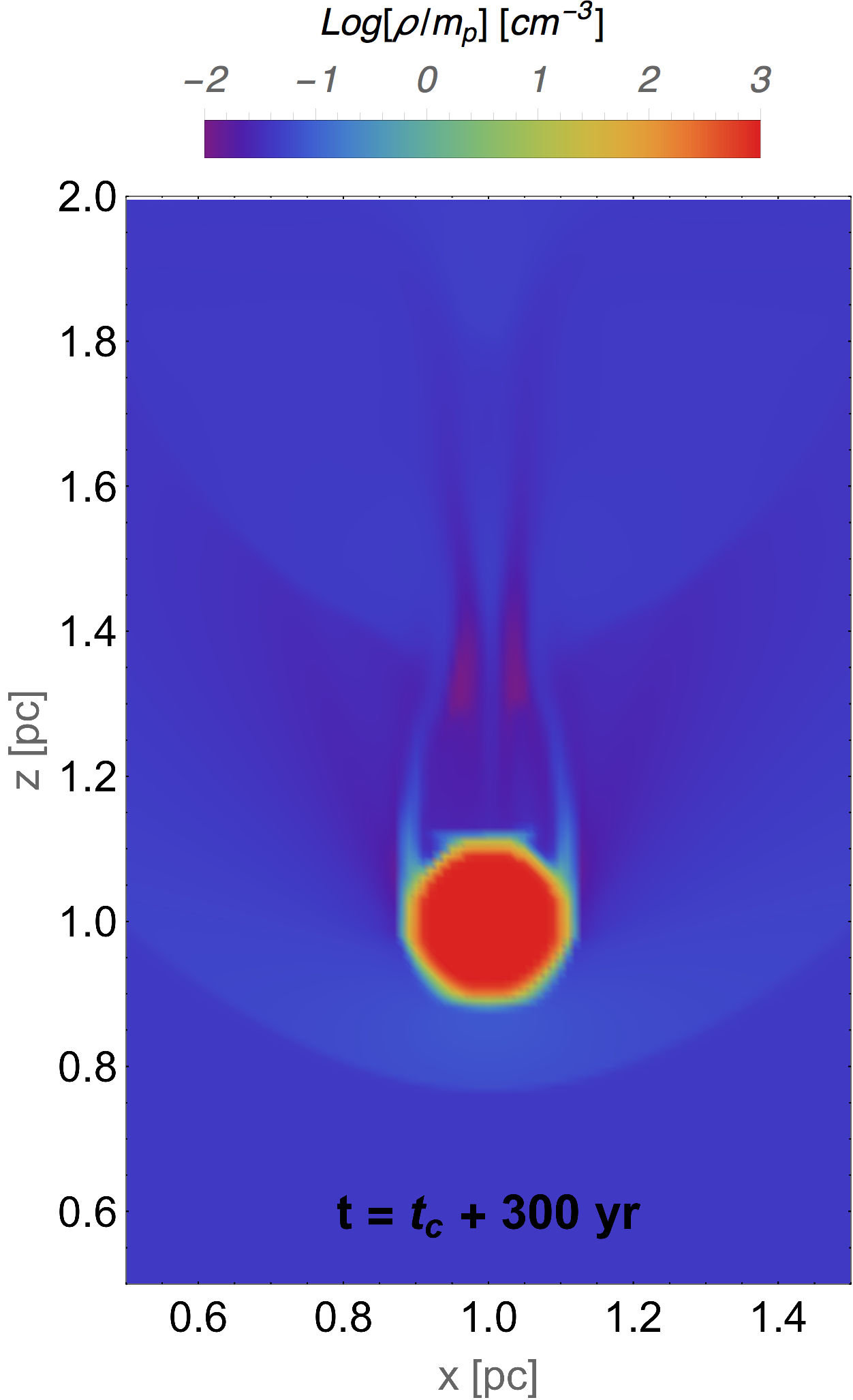}
\caption{Mass density of the plasma for the 3D MHD simulation in the oblique shock configuration with density contrast $\chi=10^5$. The plots show a 2D section along $y=1$~pc, passing through the centre of the clump. From left to right, the evolution is shown for $50$, $150$ and $300$~yr after the first shock-clump contact, occurring at $t=t_c$.}
\label{fig:rho6Obl} 
\end{figure*}

\begin{figure*}
\centering
\includegraphics[scale=0.52]{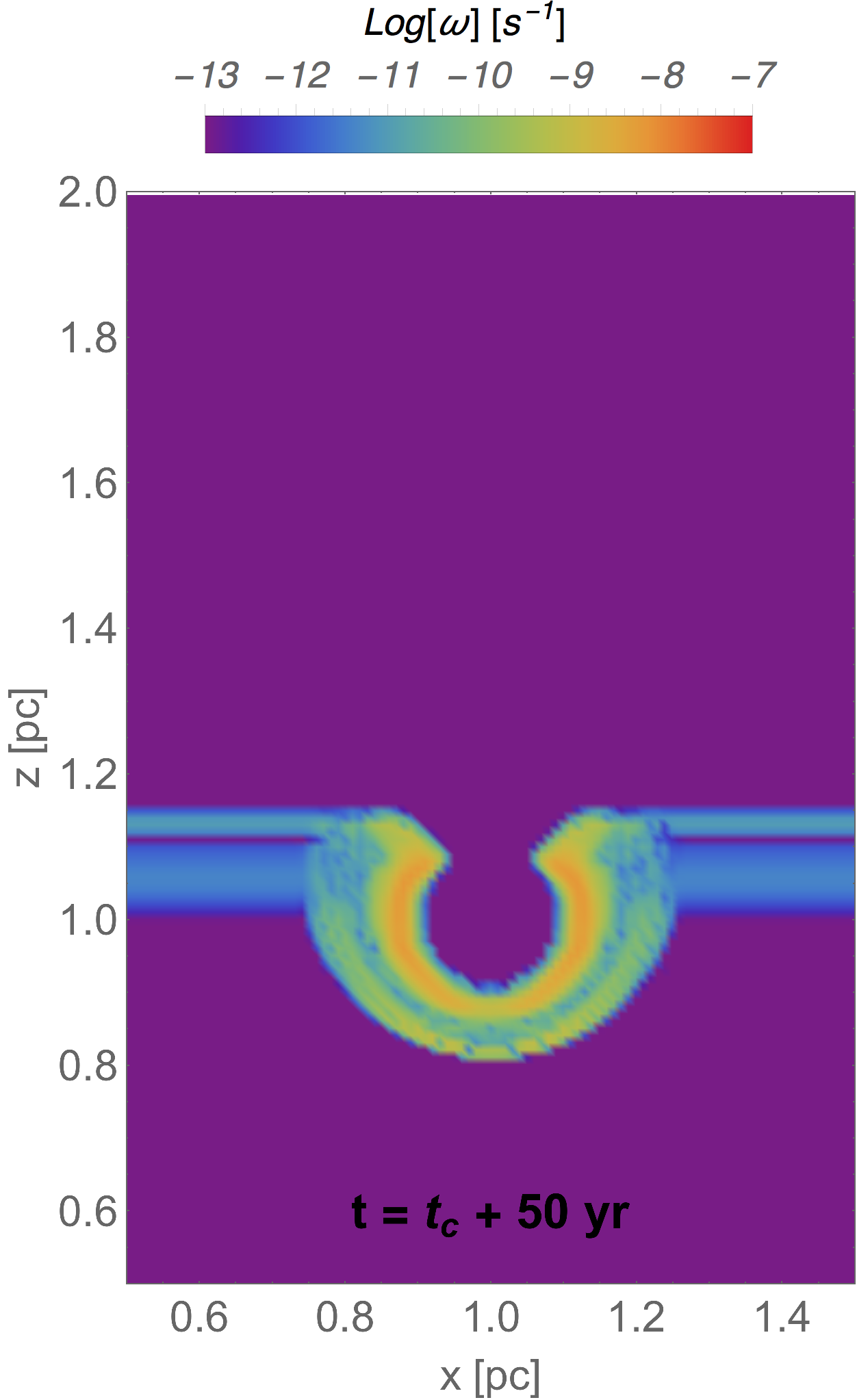}
\includegraphics[scale=0.52]{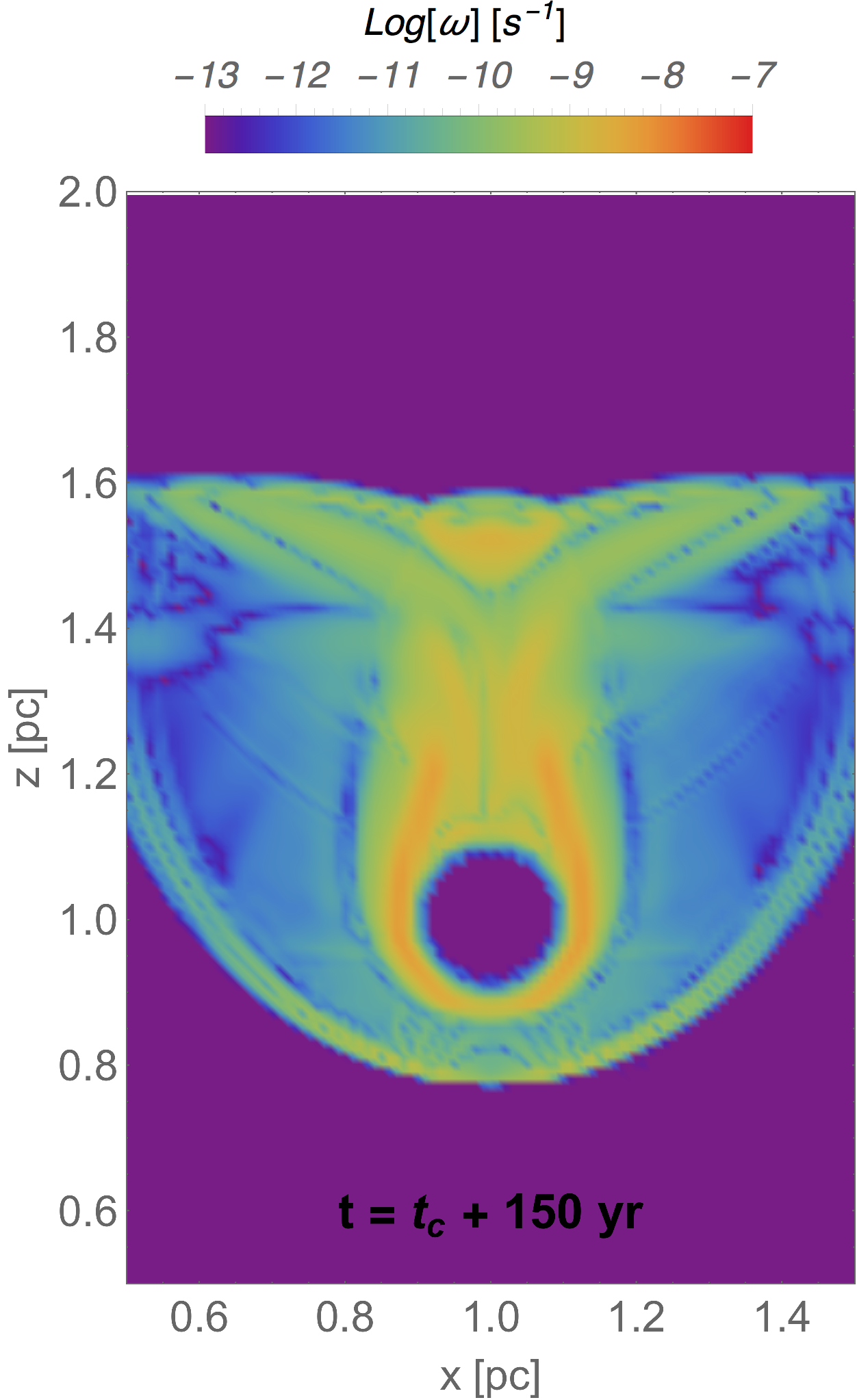}
\includegraphics[scale=0.52]{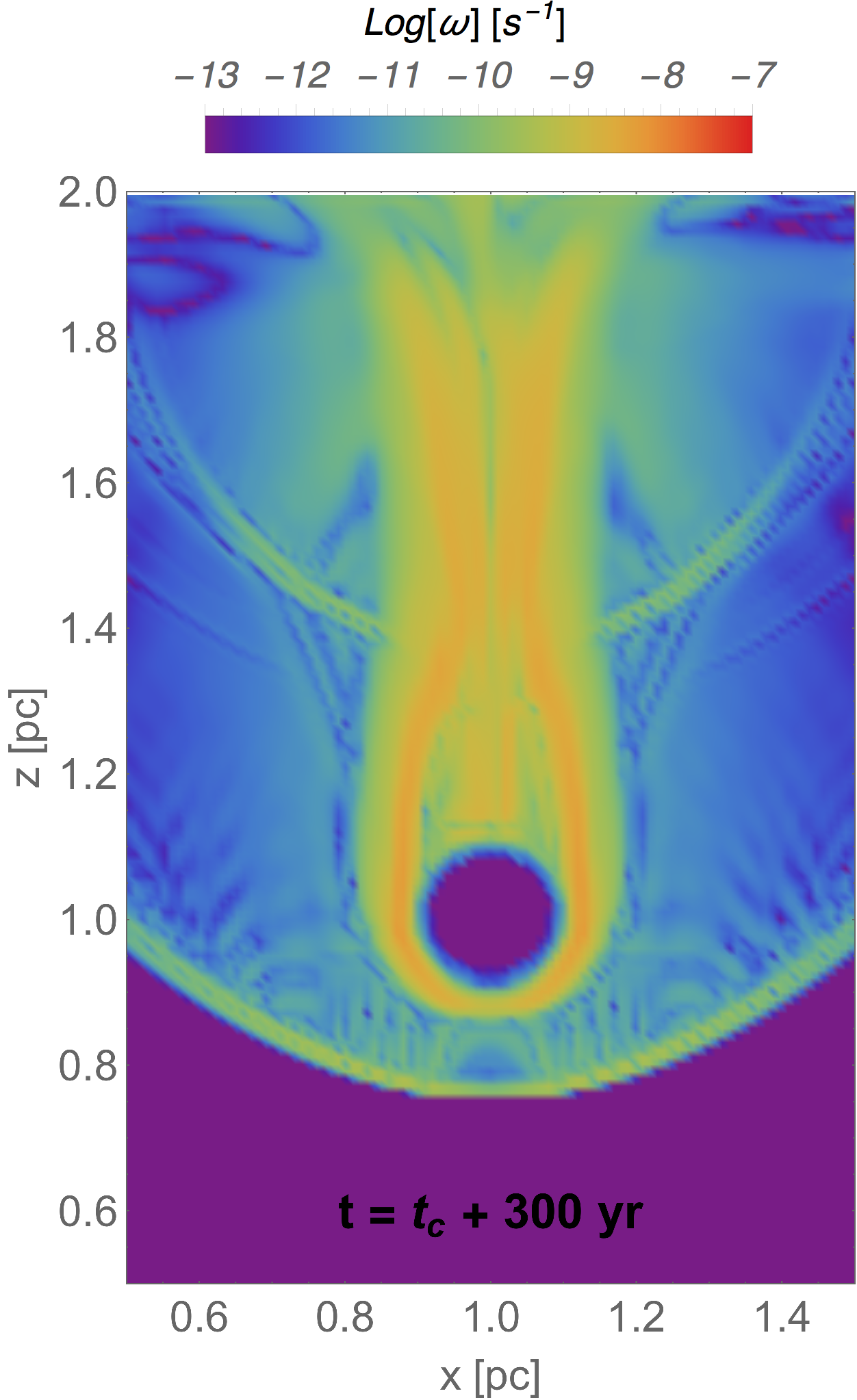}
\caption{Vorticity $\omega = \mathbf{\nabla} \times \mathbf{v}$ of the plasma for the same simulation shown in Fig.~\ref{fig:rho6Obl}.}
\label{fig:vorticityObl} 
\end{figure*}

\begin{figure*}
\centering
{\includegraphics[scale=0.52]{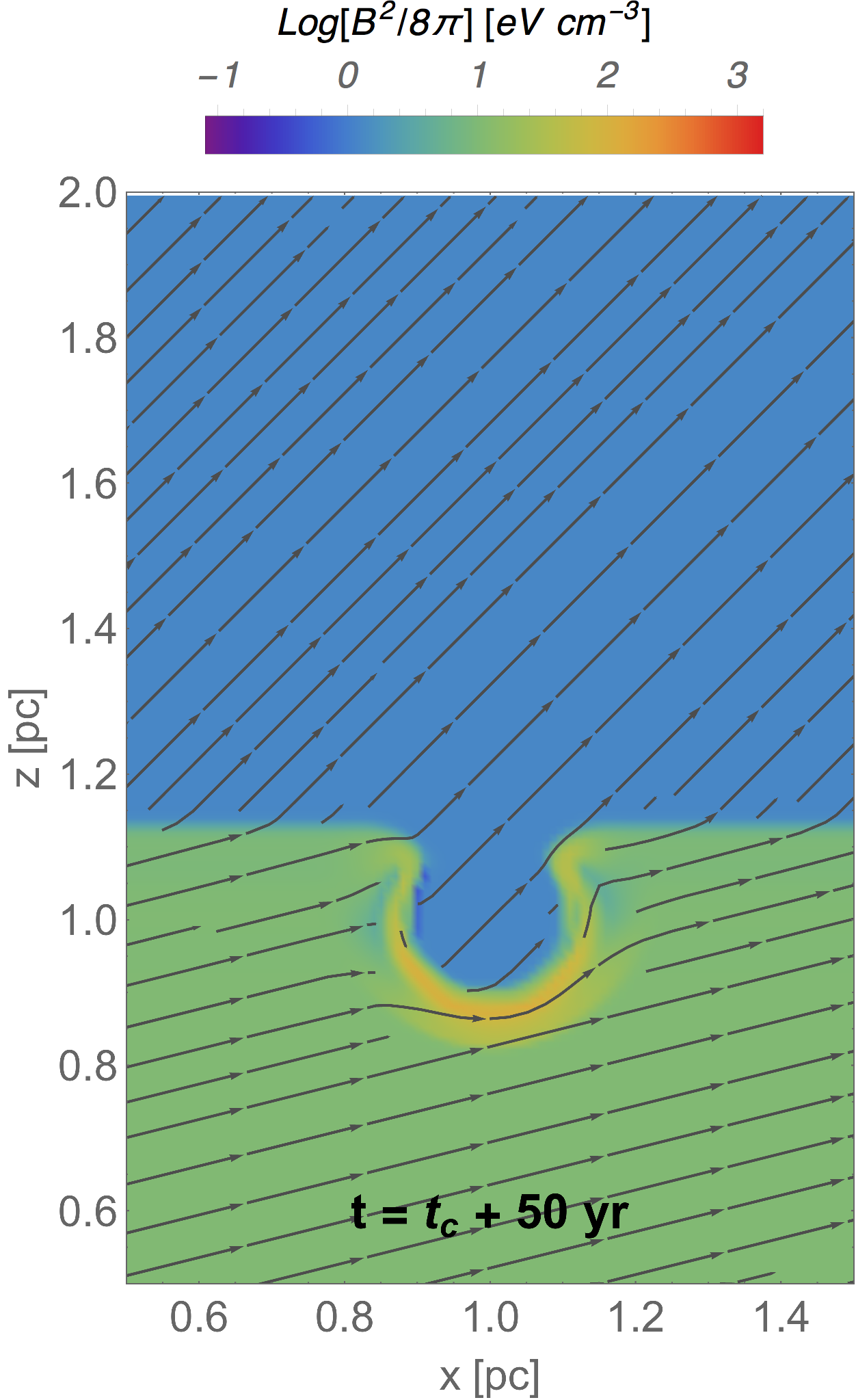}
\includegraphics[scale=0.52]{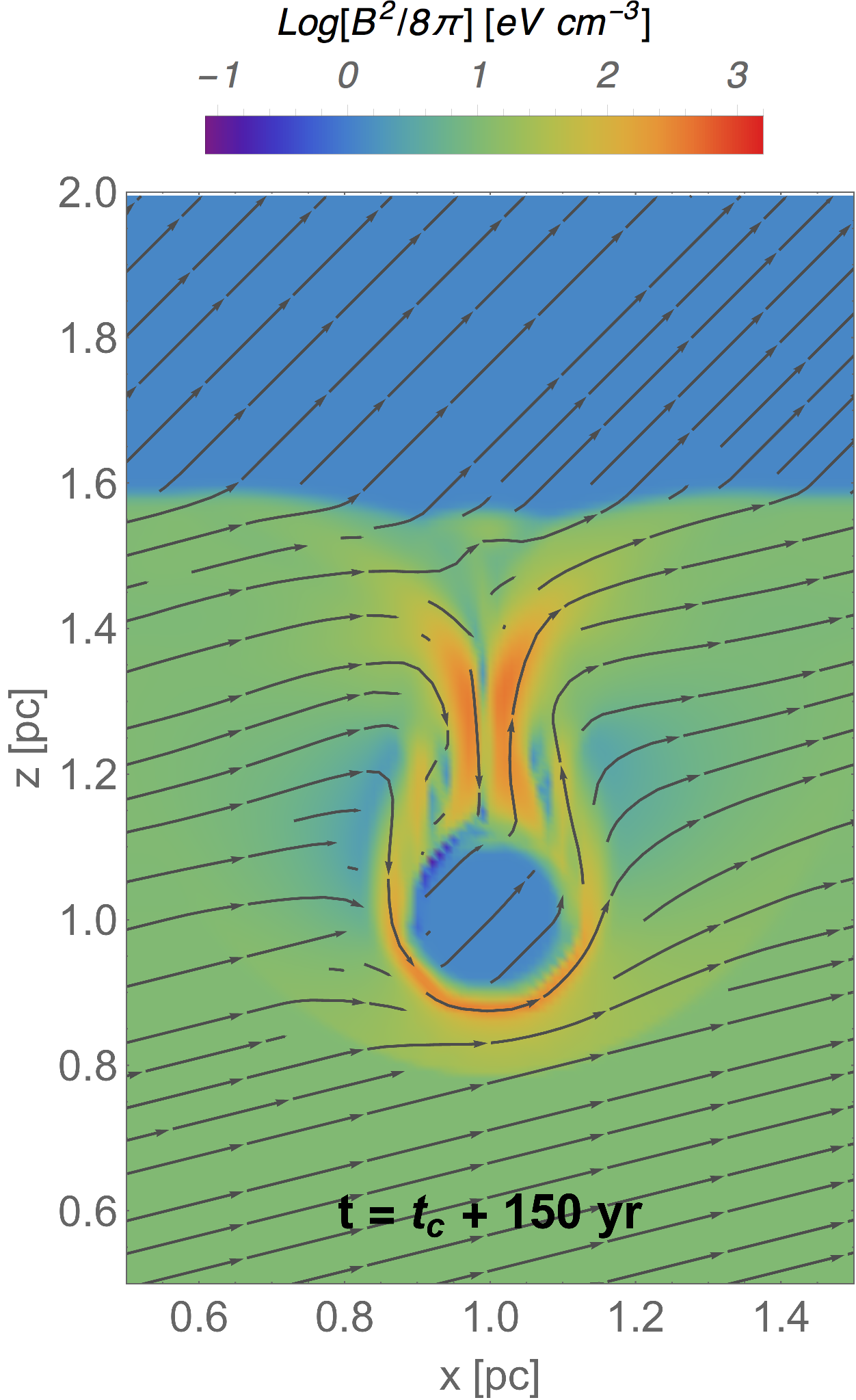}
\includegraphics[scale=0.52]{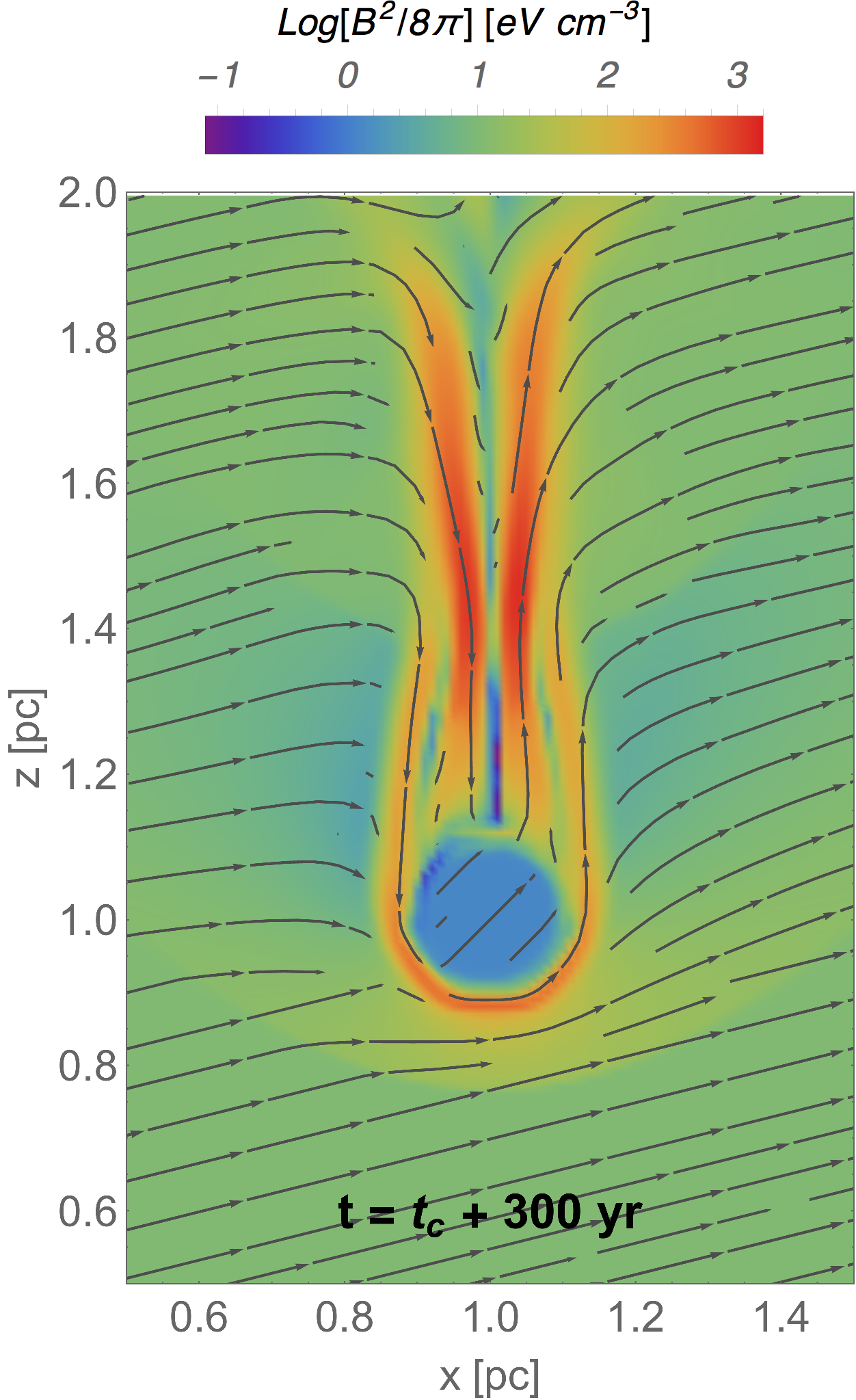}}
{\includegraphics[scale=0.52]{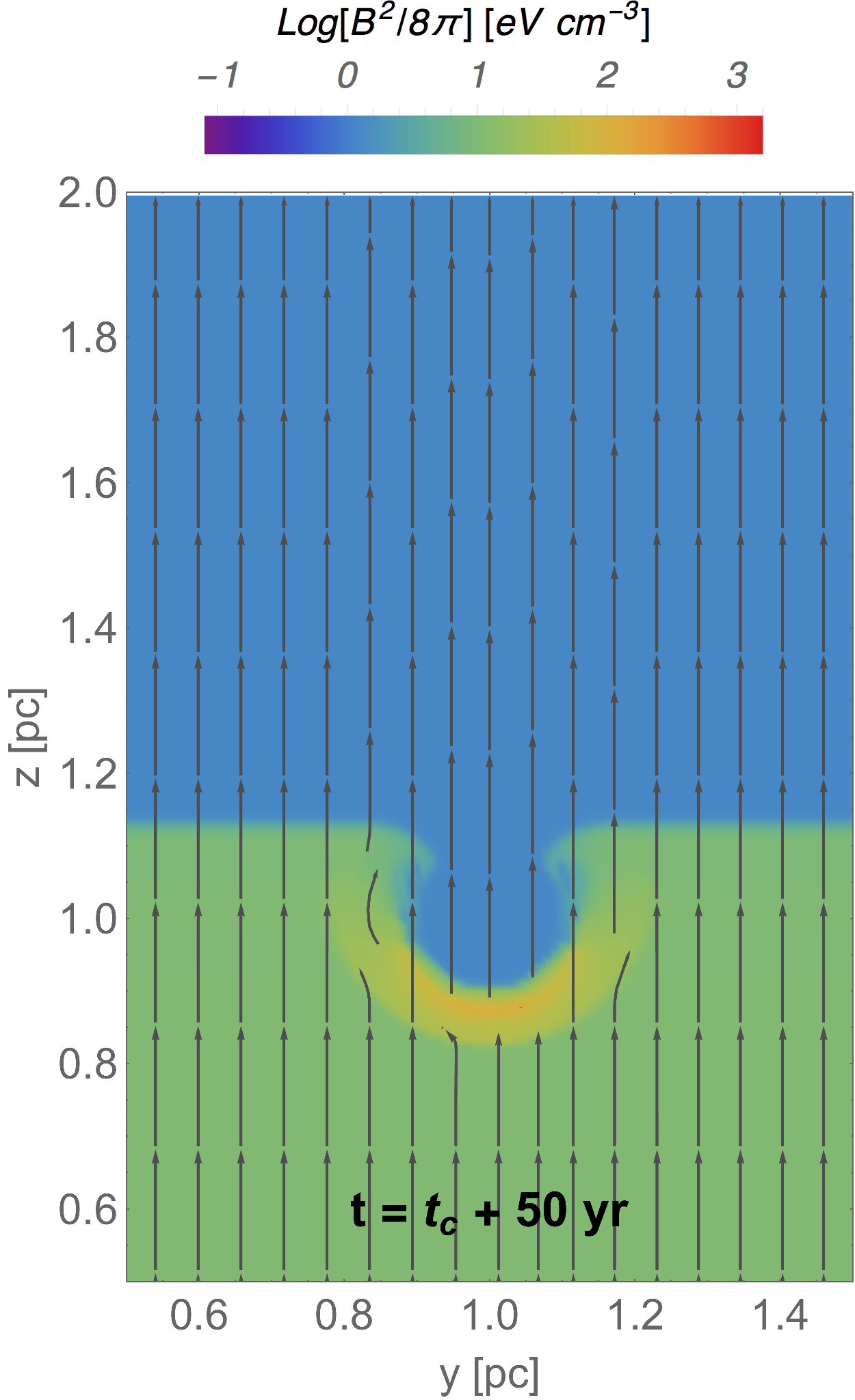}
\includegraphics[scale=0.52]{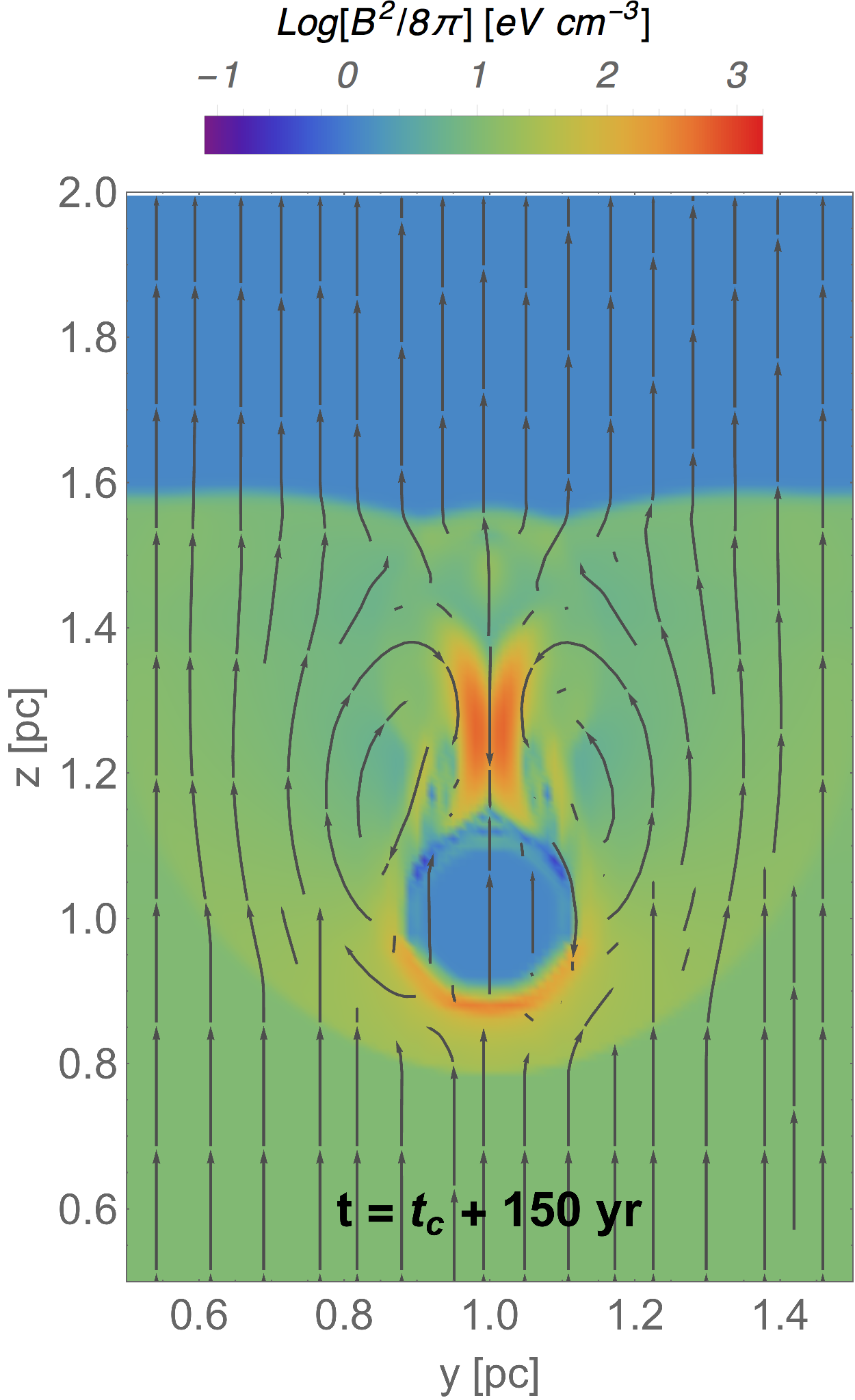}
\includegraphics[scale=0.52]{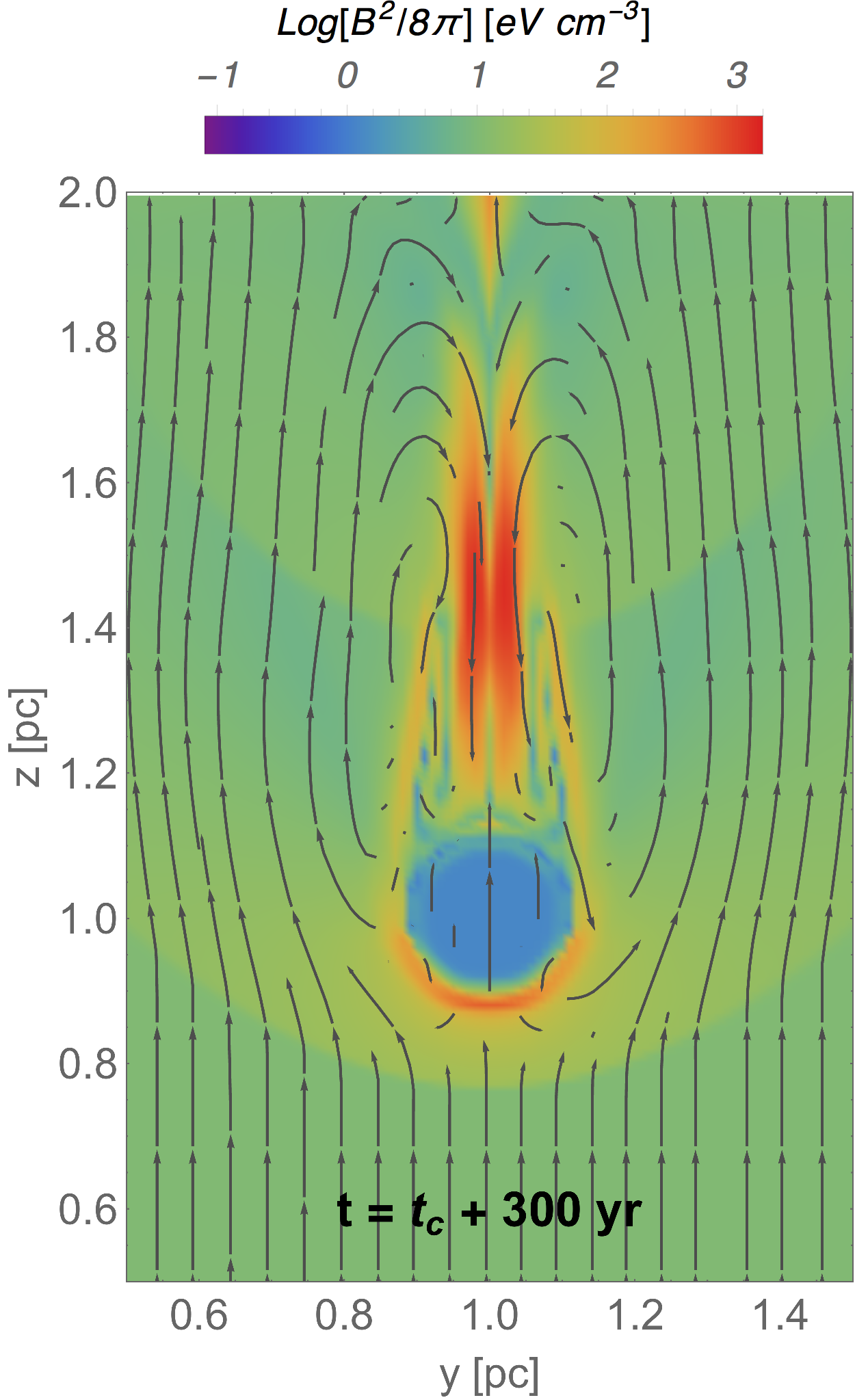}}
\caption{Energy density of magnetic field (color scale) for the same simulation shown in Fig.~\ref{fig:rho6Obl}. Upper and lower panels refer to a 2D cut across the centre of the clump in the $x-z$ and $y-z$ planes, respectively at $y=1$~pc and $x=1$~pc, and the stream lines show the direction of the regular magnetic field in the corresponding planes.}
\label{fig:MField6Obl} 
\end{figure*}

\begin{figure*}
\subfigure[]{\label{fig:v6ArObl} \includegraphics[scale=0.34]{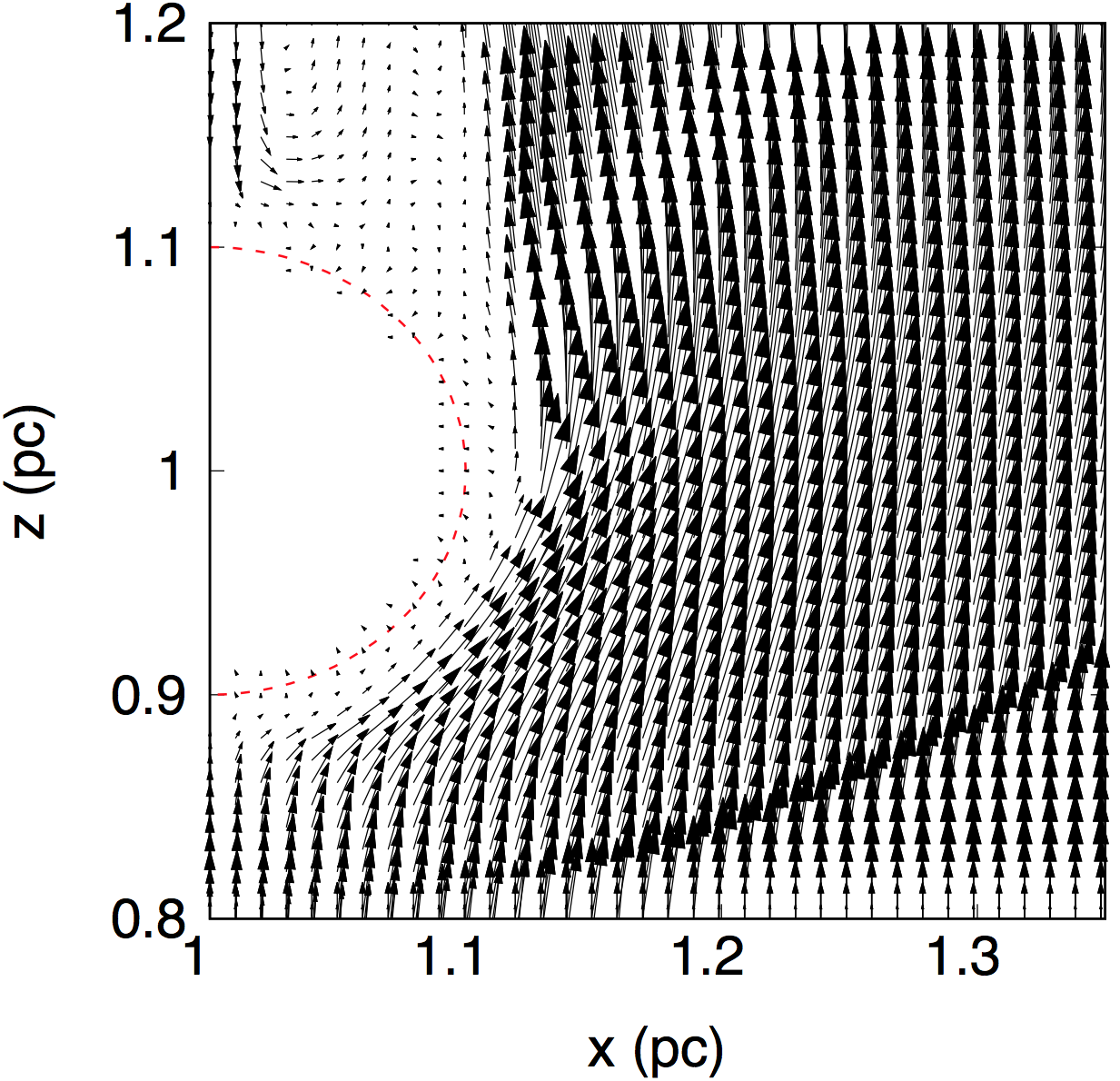}}
\subfigure[]{\label{fig:vAnalytical} \includegraphics[scale=0.4]{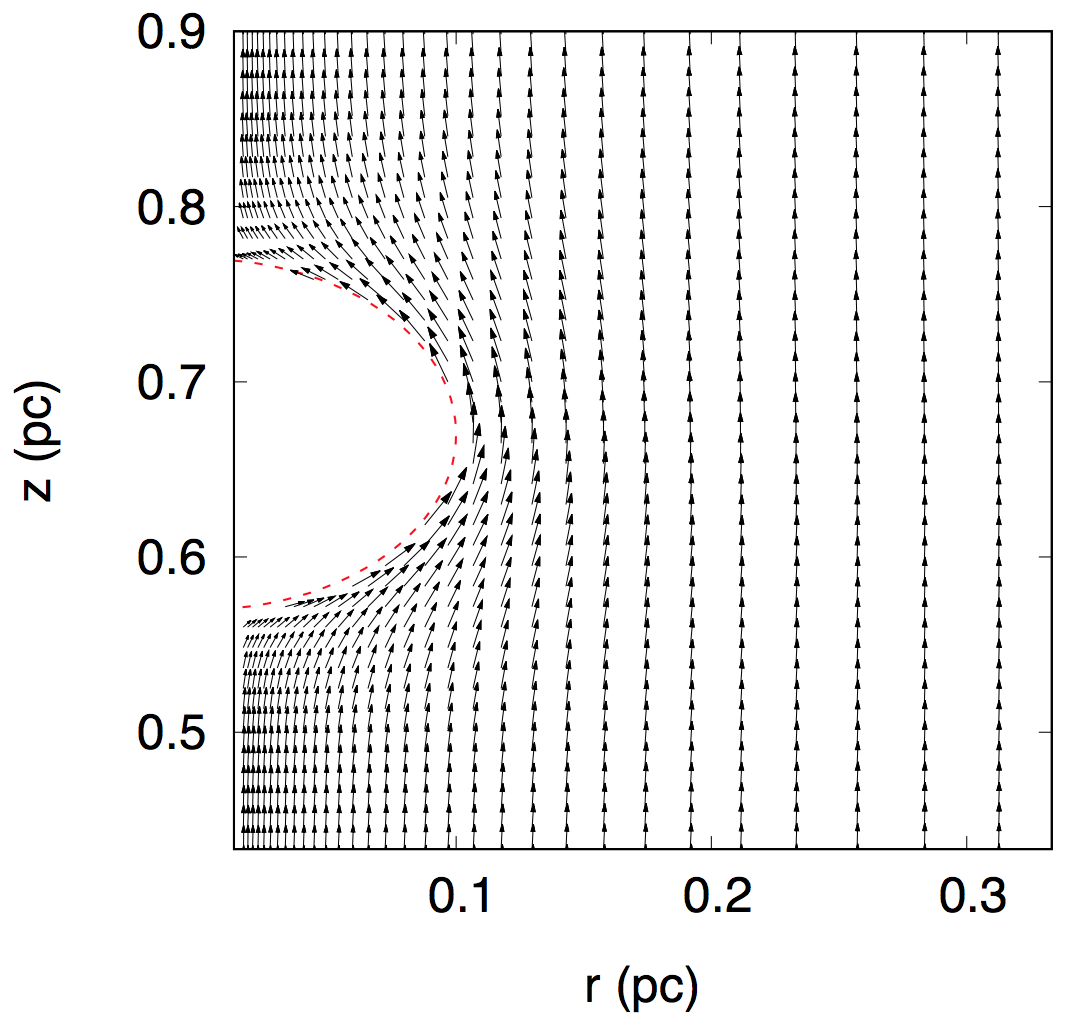}}
\caption{\textit{Left:} Velocity field resulting from the MHD simulation in the oblique shock configuration and density contrast $\chi=10^5$, at a time $t=t_c+300$~yr and in a 2D section along $y=1$~pc, passing through the center of the clump. The bow-shock produced by the clump in the shocked ISM is also visible which, however, is always very mild ($M_2 < 2$). \textit{Right:}  Analytical velocity field adopted in the numerical solution of the transport equation, fully tangential to the clump surface and directed along the $z$-direction in the far field limit. The red dashed line defines the clump size, without considering its magnetic skin.}
\label{fig:VFieldObl} 
\end{figure*}

A magnetic field of intensity $B^\textrm{up}_0=5 \mu$G is set in the region upstream of the shock. We present here a simulation of an oblique shock inclined by 45° with respect to the shock normal. We set $\mathbf{B_0}=(B_{0x},0,B_{0z})$, $B^\textrm{up}_0=5 \mu$G and $B_{0x}=B_{0z}$. We are interested in the evolution of the background plasma for about 300 years from the first shock-clump interaction, as will be described in \S~\ref{sec:gamma}. 
Within this time interval, results from the MHD simulations are shown in Figs.~\ref{fig:rho6Obl}, \ref{fig:vorticityObl}, \ref{fig:MField6Obl} and \ref{fig:VFieldObl} corresponding to plasma mass density, vorticity $\mathbf{\omega}$, magnetic energy density and plasma velocity, respectively. The plasma vorticity is here defined as $\mathbf{\omega} = \mathbf{\nabla} \times \mathbf{v}$. These results can be summarized as follows:
\begin{itemize} 
\item[I)]  The clump maintains its density contrast, although the density distribution tends to smoothens, as seen in Fig.~\ref{fig:rho6Obl};
\item[II)]  During the whole simulated time the shock has not yet crossed the clump, as represented by the plasma velocity field lines in Fig.~\ref{fig:v6ArObl};
\item[III)] The regular magnetic field is wrapped around the clump surface in a region with a typical size of $R_c/2$, as shown in Fig.~\ref{fig:MField6Obl}. Comparing Figs. \ref{fig:vorticityObl} and \ref{fig:MField6Obl}, one can see that the magnetic field in the clump skin results amplified from compression and stretching due to the strong vorticity developed in the plasma. The magnetic amplification is most effective where the vorticity is the largest, meaning that the shear amplification is more important than pure compression as already pointed out by, e.g., \cite{MacLow94} and \cite{Jones96}\footnote{In the context of the present work it is important to stress that we do not see any turbulence developing in the clump skin. Vorticity stream and magnetic field lines are regular in the sense that they do not show eddies; the vorticity observed is only due to the shear. As a consequence, our simulations do not support the idea that a turbulent cascade develops all around the clump surface. On the other hand, it is also possible that some turbulence develops on scales much smaller than our resolution scale.}. The resulting magnetic field  energy density is $\sim 100$ times larger than in the regular downstream region: the amplification factor obtained directly follows from the simulation set up, namely from the shock Mach number, which is set in order to reproduce the conditions expected to operate in RX~J1713.7-3946. The fact that the magnetic field around the clump is mainly directed along the tangential direction implies that it is difficult for accelerated particles to diffuse orthogonally to the clump surface, along the radial direction, as will also be discussed in \S~\ref{subsec:perpD}. 
\item[IV)] In the region immediately behind the clump, a long tail is formed where the plasma develops eddies and will eventually become turbulent, as observed in Fig.~\ref{fig:v6ArObl}. Also in this region, the magnetic field is amplified up to a factor $\sim 10$. However, the tale behind the clump is not particularly relevant for the CR propagation inside the clump, but it can be important when considering the synchrotron X-ray emission from electrons (see \S~\ref{subsec:xray}).
\end{itemize} 

In order to understand whether the magnetic amplification is peculiar of our chosen configuration, we also performed MHD simulations changing the initial magnetic field orientation and the density contrast. Concerning the orientation, the only case where the amplification results negligible is when the magnetic field is purely parallel to the shock normal: in this situation, the magnetic field around the clump is not compressed and also the shear is less effective, while the amplification in the tail is observed as due to plasma vorticity. To explore, instead, the effect of the density contrast we performed simulations with $\chi = 10^2$, $10^3$ and $10^4$.  In these cases, for a fixed shock speed, we expect the clump to evaporate on shorter time scales than the one given by Eq.~(\ref{eq:taucc}). We found that an effective amplification of the magnetic field around the clump is obtained if $\chi \gtrsim10^3$. For less massive clumps, where the evaporation time is comparable to the remnant age, two differences arise: i) once heated, these clumps would contribute to the thermal emission of the remnant and ii) the resulting gamma-ray spectrum would not manifest a pronounced hardening. 
Hence, detailed spectroscopic and morphological observations are crucial for providing a lower limit on the density contrast of the circumstellar medium (CSM). A scenario where the SNR shock might be accelerating particles into a cloudy ISM with a density contrast as low as $\chi = 10^2$ was considered in \citet{berezko} to explain the soft spectrum observed in the Tycho's SNR.

\section{Particle transport}
\label{sec:transport}

\subsection{Transport equation} 

CRs are scattered by MHD waves parallel to the background magnetic field. The transport equation which describes the temporal and spatial evolution of the CR density function in the phase space $f(\mathbf{x},\mathbf{p},t)$ reads as (\citealp{ginzburg} and \citealp{drury})
\begin{equation}
\label{eq:transport}
\frac{\partial f}{\partial t} + \mathbf{v} \cdot \mathbf{\nabla} f= \mathbf{\nabla} \cdot \left[ D \, \mathbf{\nabla} f \right] + \frac{1}{3}p\frac{\partial f}{\partial p} \mathbf{\nabla} \cdot \mathbf{v} + Q_\textrm{CR}
\end{equation}
\noindent
where $D$ is the diffusion coefficient, $p$ is the particle momentum and $Q_\textrm{CR}=f_\textrm{inj}v_s \delta(z-z_s)$ is the injection flux, assumed to take place at the shock, namely in $z=z_s$. Advection in the background velocity field is expressed by the left hand side of Eq.~(\ref{eq:transport}), together with the time variation of the particle density function, while diffusion, plasma adiabatic compression and injection are expressed on the right hand side. Given the system symmetry, we will solve this equation in cylindrical coordinates, through a finite difference method. A grid of $2$~pc$\times 2$~pc is set, with a spherical clump located at $(r_0,z_0)=(0,0.67)$~pc. A logarithmic step is used in the radial dimension, while a uniform spacing is fixed along the shock direction. The spatial resolution of the grid is set in such a way that, for each simulated momentum, the proton energy spectrum reaches a convergence level better than 5\%. We set an operator splitting scheme based on an the Alternated Direction Implicit (ADI) method, flux conservative and upwind, second order in both time and space, subject to a Courant condition $C = 0.8$. The initial condition of the simulation includes the presence of a shock precursor in the upstream region (everywhere but inside the clump, which starts empty of CRs): it represents the equilibrium solution of the diffusive-advective transport equation, such that
\begin{equation}
\label{eq:ci}
f(r,z,p,t=0)=f_0(p) \exp \left[ -\frac{(z-z_s) v_s}{D(p)} \right]
\end{equation}
\noindent
Here $f_0(p)$ represents the particle spectrum at the shock location $z_s$. We set a $p^{-4}$ power-law spectrum, following the test-particle approach of Diffusive Shock Acceleration (DSA) theory (see \citealp{skilling, bell, blanford}). We added an exponential cut-off in momentum $p_\textrm{cut}$ in order to take into account the maximum attainable energy, resulting in
\begin{equation}
\label{eq:dsa}
f_0(p) \propto p^{-4} \exp \left[-\frac{p}{p_\textrm{cut}} \right]
\end{equation}
\noindent
Boundary conditions are such that a null diffusive flux is set on every boundary, except in the upstream of the $z$-direction, where the precursor shape is set. 

When the forward shock hits the clump, we assume that the transmitted shock does not accelerate particles because it is very slow and it is propagating in a highly neutral medium. Moreover, we neglect the reflected shock which propagates back into the remnant, since it would not contribute to the acceleration of particles because of its low Mach number $M_2 \lesssim 2$. Given the result shown in \S~\ref{subsec:mhd}, we set an analytical velocity field, irrotational and divergence-less through all the space (except at the shock and clump surface). This is obtained by solving the Laplace equation in cylindrical coordinates for the velocity potential, with the boundary conditions that in the far field limit the downstream velocity field is directed along the shock direction and equal to $\mathbf{v}=v_\textrm{down} \widehat{z} = \frac{3}{4} v_s \widehat{z}$, while at the clump surface the field is fully tangential. The resulting solution reads as
\begin{equation}
\begin{dcases}
v_r (r,z) & =-\frac{3}{2} \frac{R^3_c r z}{(r^2+z^2)^{5/2}} v_\textrm{down} \\
v_z (r,z) & = \left[ 1+ \frac{1}{2} R^3_c \frac{(r^2-2z^2)}{(r^2+z^2)^{5/2}} \right] v_\textrm{down}
\end{dcases}
\label{eq:vfield}
\end{equation}
With such a choice of the velocity field, the adiabatic compression term vanishes. Moreover, we set a null velocity field inside the clump, since $v_{s,c} \ll v_s$, and in the upstream region. A schematic view of the velocity vector field adopted is given in Fig.~\ref{fig:vAnalytical}. Comparing it with the results from MHD simulation obtained in \S~\ref{subsec:mhd} and showed in Fig.~\ref{fig:v6ArObl}, one can see that the two velocity profiles are quite similar except for the turbulent region behind the clump.

Furthermore, we will consider a stationary space-dependent and isotropic Bohm diffusion through all the space, so that 
\begin{equation} 
\label{eq:Bohm}
D_\textrm{Bohm}(\mathbf{x},p)=\frac{1}{3} r_L(\mathbf{x},p) v(p) = \frac{1}{3} \frac{pc}{ZeB_0 (\mathbf{x})} v(p)
\end{equation}
\noindent
where $r_L(\mathbf{x},p)$ is the Larmor radius of a particle with charge $Ze$ in a background magnetic field $B_0 (\mathbf{x})$. In the following we will only consider relativistic protons, with momenta $p \in [1 \, \textrm{GeV/c} - 1 \, \textrm{PeV/c}]$, since this is the energy interval relevant for the production of high-energy and very-high-energy gamma rays.  We set a space-dependent magnetic field $B_0(\mathbf{x})$, defining four regions in the space: 
\begin{itemize} 
\item[I)] The unshocked CSM;
\item[II)] The shocked CSM, where $B_0 \equiv B_\textrm{CSM} = 10 \, \mu$G;
\item[III)]  The clump interior, with a size of $R_\textrm{c}=0.1$~pc, where diffusion is not efficient such that $B_0 \equiv B_c = 1 \, \mu$G;
\item[IV)]  The clump skin, with a size of $R_\textrm{s}=0.5 R_\textrm{c}$ around the clump itself, where the amplification of the magnetic field is realized such that $B_0 \equiv B_s = 100 \, \mu$G.
\end{itemize} 

\begin{figure}
\centering
\subfigure[]{\label{fig:z1} \includegraphics[scale=0.3]{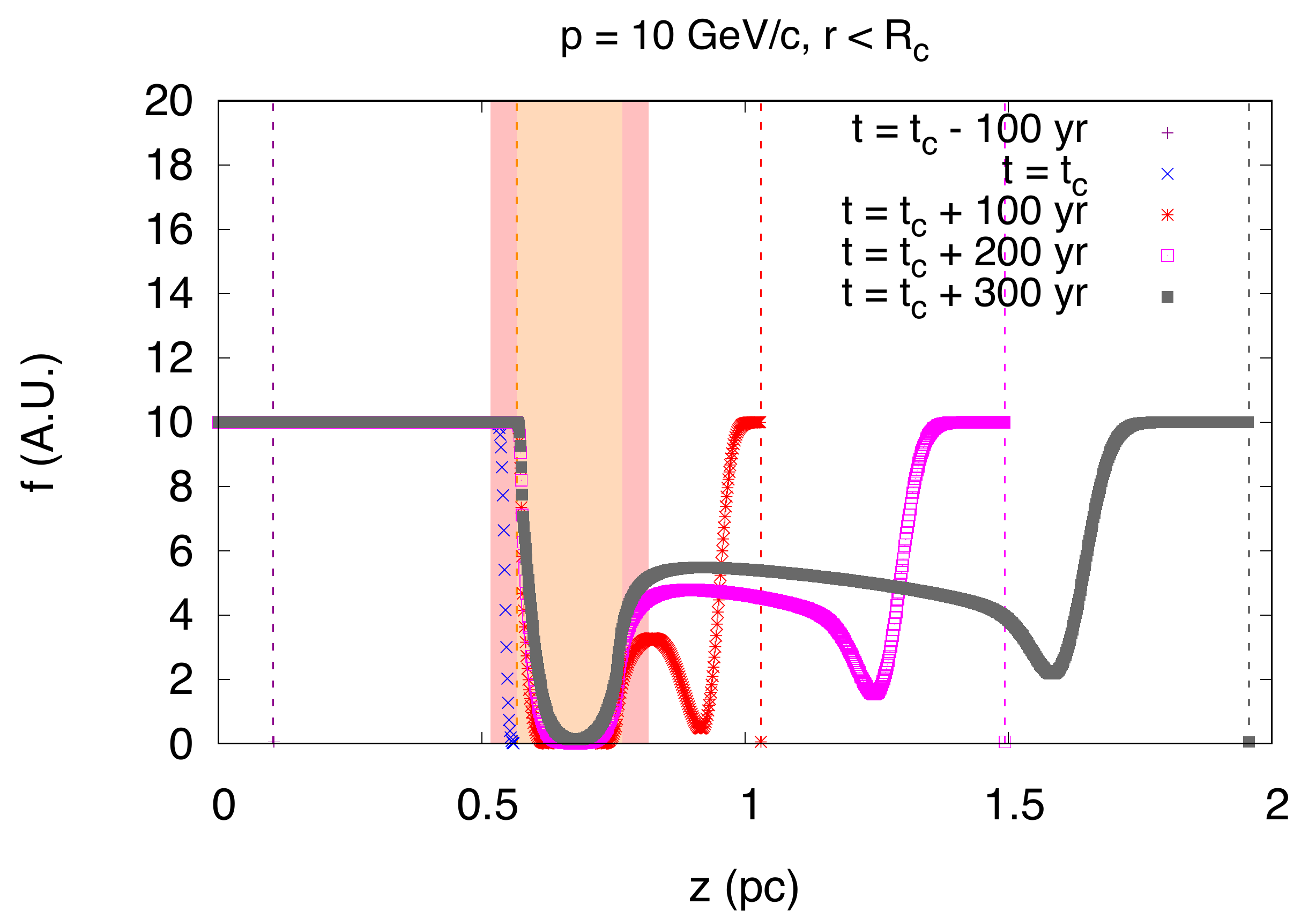}}
\subfigure[]{\label{fig:z3} \includegraphics[scale=0.3]{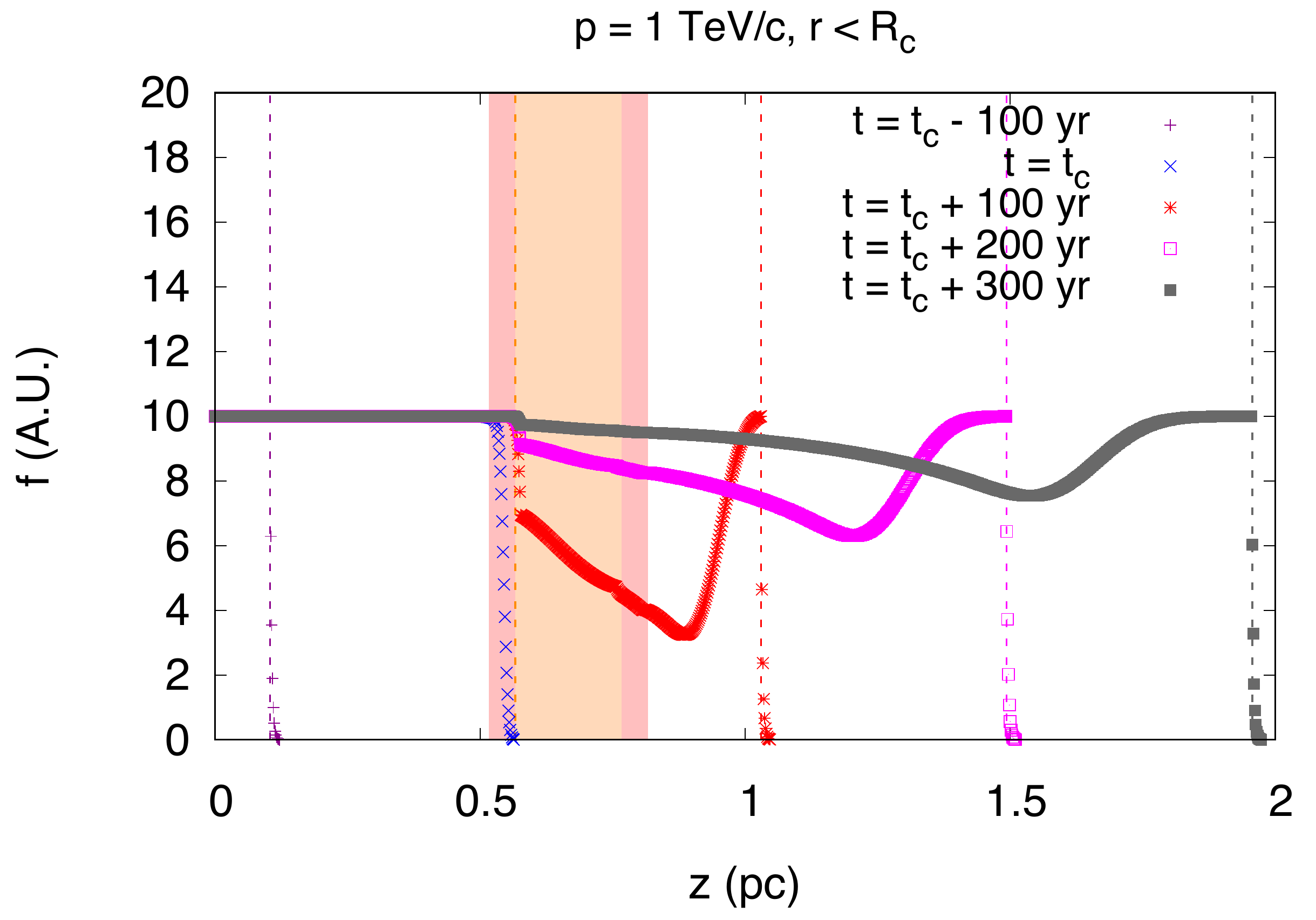}}
\subfigure[]{\label{fig:z5} \includegraphics[scale=0.3]{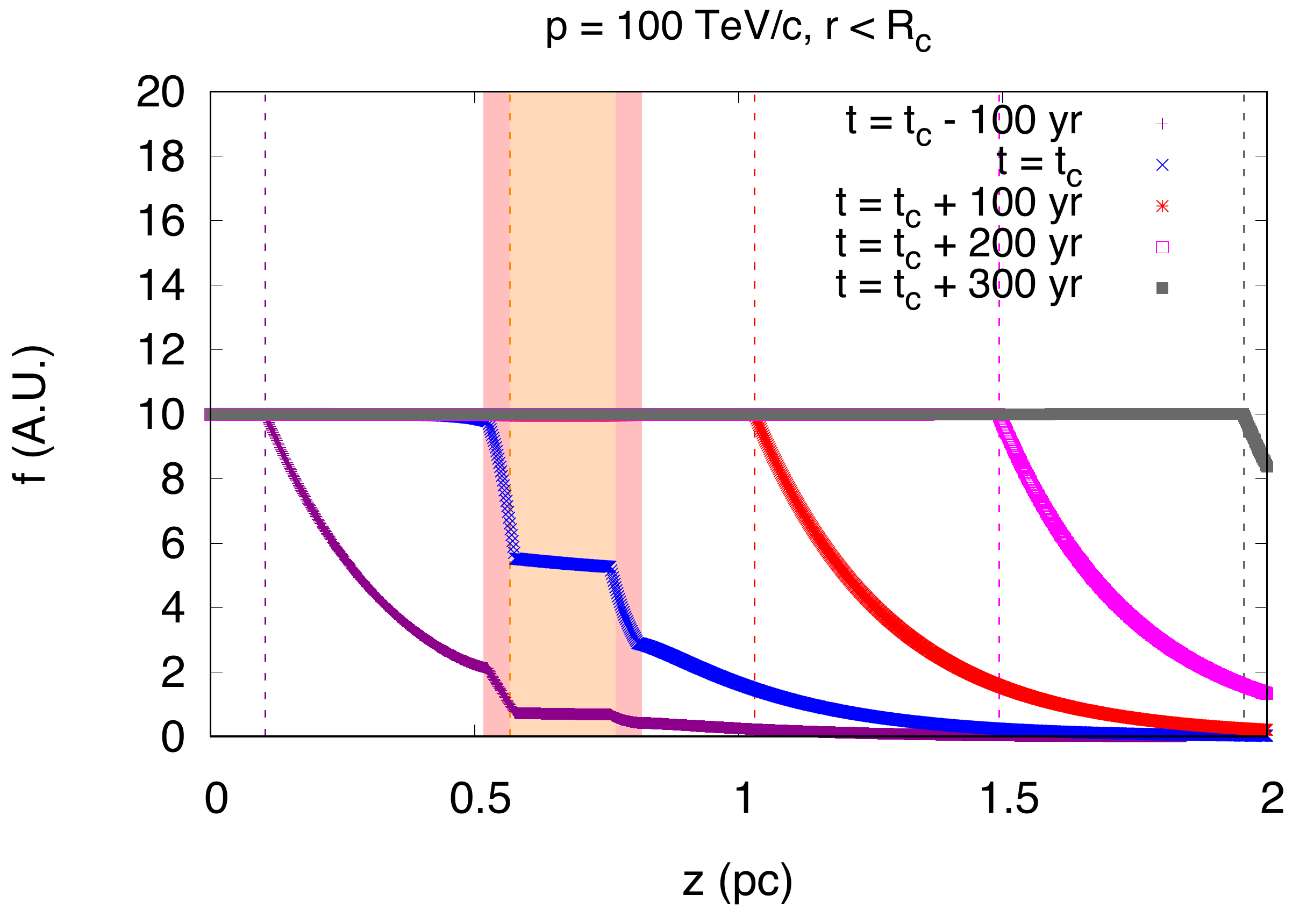}}
\caption{Density profiles of the accelerated particle in the shock region along $z$ (at fixed radial position inside the clump) for CR protons of momentum (a) $10$~GeV/c, (b) $1$~TeV/c and (c) $100$~TeV/c. Vertical dashed lines represent the shock position at a given time, as indicated in the legend. The light pink band defines the clump interior, while the dark pink band defines its magnetic skin.}
\end{figure}

\begin{figure*}
\centering
\subfigure[]{\label{fig:grid3_1} \includegraphics[scale=0.5]{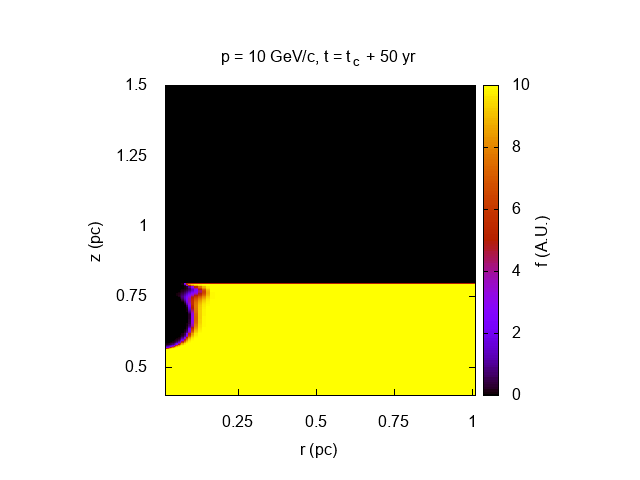}}
\subfigure[]{\label{fig:grid3_4} \includegraphics[scale=0.5]{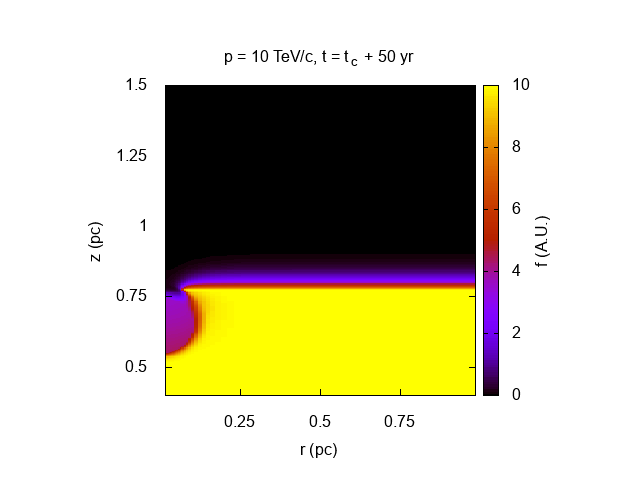}}
\subfigure[]{\label{fig:grid4_1} \includegraphics[scale=0.5]{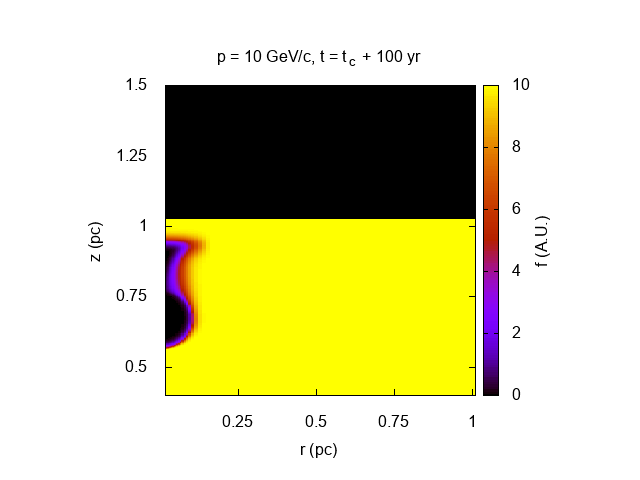}}
\subfigure[]{\label{fig:grid4_4} \includegraphics[scale=0.5]{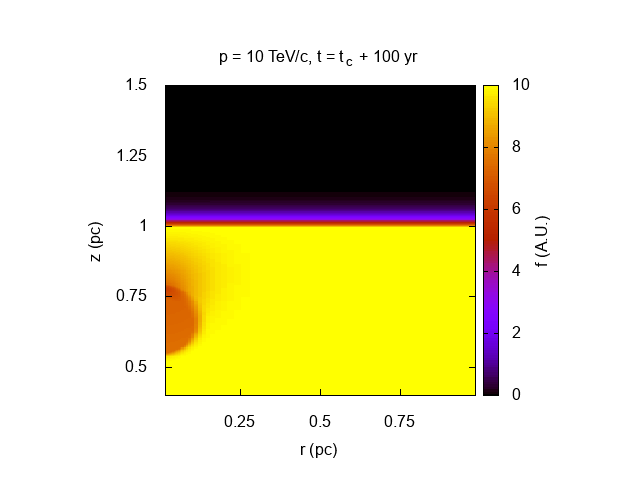}}
\subfigure[]{\label{fig:grid6_1} \includegraphics[scale=0.5]{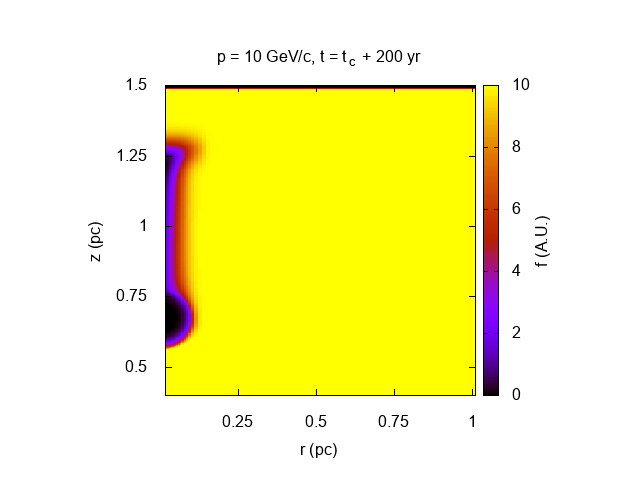}}
\subfigure[]{\label{fig:grid6_4} \includegraphics[scale=0.5]{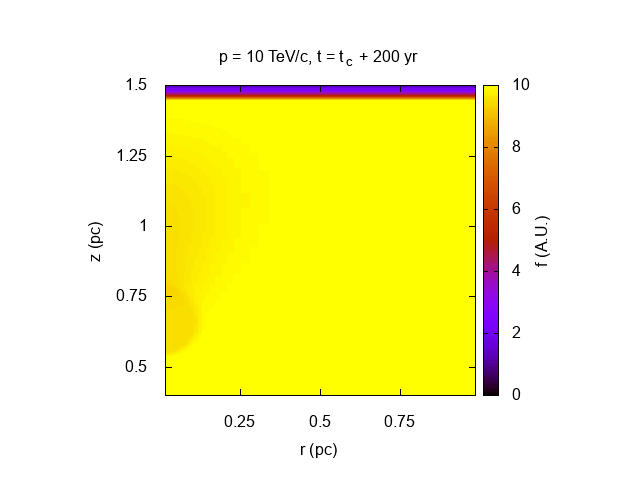}}
\caption{Distribution function of CR protons of momentum (a,c,e) 10 GeV/c and (b,d,f) 10~TeV/c at different times with respect to $t_c$: panel (a) and (b) are at $t=t_c+50$~yr, (c) and (d) are at $t=t_c+100$~yr, while (e) and (f) are at $t=t_c+200$~yr. The precursor presence in front of the shock is well visible at 10~TeV.}
\label{fig:grids}
\end{figure*}

The density profile of accelerated particles diffusing in the region of interaction between the shock and the clump is shown in Figs.~\ref{fig:z1}, \ref{fig:z3} and \ref{fig:z5} for particles of different energy. The distribution function is flat in the downstream region, while a precursor starts at the shock position, as defined in Eq.~(\ref{eq:ci}). 
As expected, low-energy particles penetrate inside the clump at much later times with respect to high-energy ones. The time evolution of the CR distribution function for different CR momenta is shown in Fig.~\ref{fig:grids}: Figs.~\ref{fig:grid3_1}, \ref{fig:grid4_1} and \ref{fig:grid6_1} refer to 10~GeV particles, while Figs.~\ref{fig:grid3_4}, \ref{fig:grid4_4} and \ref{fig:grid6_4} refer to 10~TeV particles. It is apparent that low-energy particles are not able to fill uniformly the clump interior on the temporal scale relevant for the gamma-ray emission  (around few hundred years, as explained in \S~\ref{sec:gamma}).

Note that the value of $B_c$ used above is smaller than the strength of the large scale magnetic field expected in molecular clouds. We chose such a smaller value as representative of the effective turbulent magnetic field, which determines the diffusion coefficient in Eq.~(\ref{eq:Bohm}), and is damped by the presence of the ion-neutral friction (see \S~\ref{subsec:streaming}).

The diffusion coefficient in the shock region should be close to the Bohm regime in order to obtain an effective acceleration of protons to multi-TeV energies. For this reason in Eq.~(\ref{eq:Bohm}) we assumed isotropic diffusion with $\delta B \simeq B_0$. From a theoretical point of view, such a small diffusion is justified by the self-generation of waves from accelerated particles. Nevertheless the correct description of the diffusion in the skin region of the clump is not a trivial task. In particular it is not guaranteed that the turbulent component of the field is also amplified at a level such that the Bohm diffusion still applies. In resolving the transport equation, Eq.~(\ref{eq:transport}), we assume that this is indeed the case, namely $\delta B_s \simeq B_s$. 
Nevertheless, if the amplification of the turbulence does not occur at the same level, and $\delta B_s < B_s$, the large scale magnetic field dominates and we can envision two opposite situations: {\it a}) the case where magnetic field lines penetrate inside the clump (which occurs in a small portion of the clump surface located in the back of the cloud, where the magnetic tail develops, as shown in Fig.~\ref{fig:MField6Obl}), and {\it b}) the case where these lines stay parallel to the clump surface. While in the former case the relevant diffusion coefficient for particles to penetrate the clump is the one parallel to the magnetic field lines, in the latter we need to account for the perpendicular diffusion as well. In \S~\ref{subsec:streaming} and \ref{subsec:perpD} we discuss these two situations separately, showing that in both cases the effective $D$ is reduced with respect to the plain downstream.

\subsection{Growth and damping of MHD waves}
\label{subsec:streaming}
In the following, we investigate the effect of the streaming instability on the amplification of the turbulent magnetic field, as it might significantly contribute to the CR scattering, especially along the magnetic field lines penetrating the clump. In the framework of non-linear theory of DSA, in fact cosmic rays themselves generate the MHD waves which they need to scatter off from one side to the other of the shock surface. If this process happens in resonant conditions, a CR particle of momentum $p$ is able to excite only magnetic waves of wavenumber $k_\textrm{res}= 1/r_L (p)$. The wave growth is then due to streaming of CRs. Thus the CR density, obtained as a solution of Eq.~(\ref{eq:transport}), affects the amount of turbulence that is generated, which in turn modifies the diffusion properties of the system as \citep{skilling71}
\begin{equation}
\label{eq:D}
D(\mathbf{x},p,t)=\frac{1}{3} r_L(p) v(p) \frac{1}{\mathcal{F}(k_\textrm{res},\mathbf{x},t)}
\end{equation}
where $\mathcal{F}(k,\mathbf{x},t)$ is the turbulent magnetic energy density per unit logarithmic bandwidth of waves with wavenumber $k$, normalized to the background magnetic energy density as
\begin{equation}
\label{eq:f}
\left( \frac{\delta B(\mathbf{x},t)}{B_0} \right)^2 = \int  \mathcal{F}(k,\mathbf{x},t) d \ln k
\end{equation}
Given the strong non linearity of the problem, it is computationally prohibitive to solve in a self-consistent way the system composed by the transport equation and by the time evolution of the wave power density, which satisfies the following equation
\begin{equation}
\label{eq:dfdt}
\frac{\partial \mathcal{F}}{\partial t} + \mathbf{v_A} \cdot  \mathbf{\nabla{\mathcal{F}}} = (\Gamma_\textrm{CR} - \Gamma_\textrm{D}) \mathcal{F}
\end{equation}
where $\Gamma_\textrm{CR}$ is the growth rate of MHD waves, $\Gamma_\textrm{D}$ is the damping rate and $v_A$ is the Alfven speed equal to
\begin{equation}
\label{eq:va}
v_A=\frac{B_0}{\sqrt{4\pi n_i m_i}}
\end{equation}
We will, instead, solve Eq.~(\ref{eq:transport}) in a stationary given magnetic field and, once the $f$ is known, we will evaluate \textit{a posteriori} the contribution of the growth and damping of resonant waves due to CR streaming.

The growth rate of the streaming instability strongly depends on the CR density gradient. We therefore expect it to be more pronounced in the clump skin, where magnetic field amplification makes diffusion very efficient, thus increasing the CR confinement time in this region. The rate can be expressed as \citep{skilling71}
\begin{equation}
\label{eq:stream}
\Gamma_\textrm{CR} (k) = \frac{16}{3} \pi^2 \frac{v_A}{B_0^2 \mathcal{F}(k)} \left[ p^4 v(p) \mathbf{\nabla} f \right]_{p=p_\textrm{res}}
\end{equation}
where $p_\textrm{res}$ represents the resonant momentum.
Amplified magnetic field can in turn be damped by non-linear damping (NLD) due to wave-wave interactions or by ion neutral damping (IND) due to momentum exchange between ions and neutrals as a consequence of the charge exchange process. Since we deal with non isolated clumps, we should also account for the typical timescale of the system. Indeed if the age of the clump is shorter than the time for damping to be effective, then waves can grow freely for a timescale equal to the clump age.  

\begin{figure*}
\centering
\subfigure[]{\label{fig:nll} \includegraphics[scale=0.33]{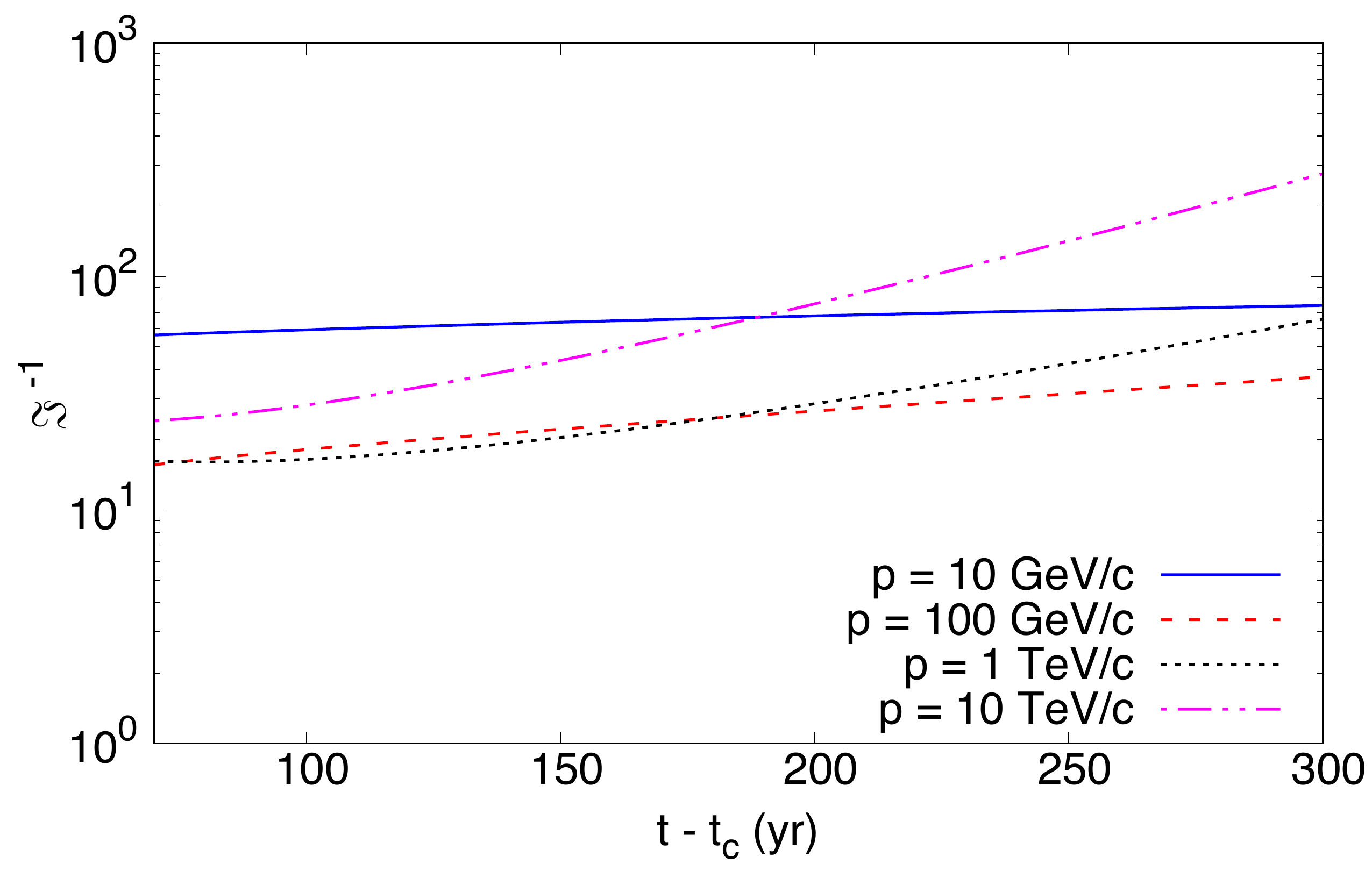}}
\subfigure[]{\label{fig:ind} \includegraphics[scale=0.33]{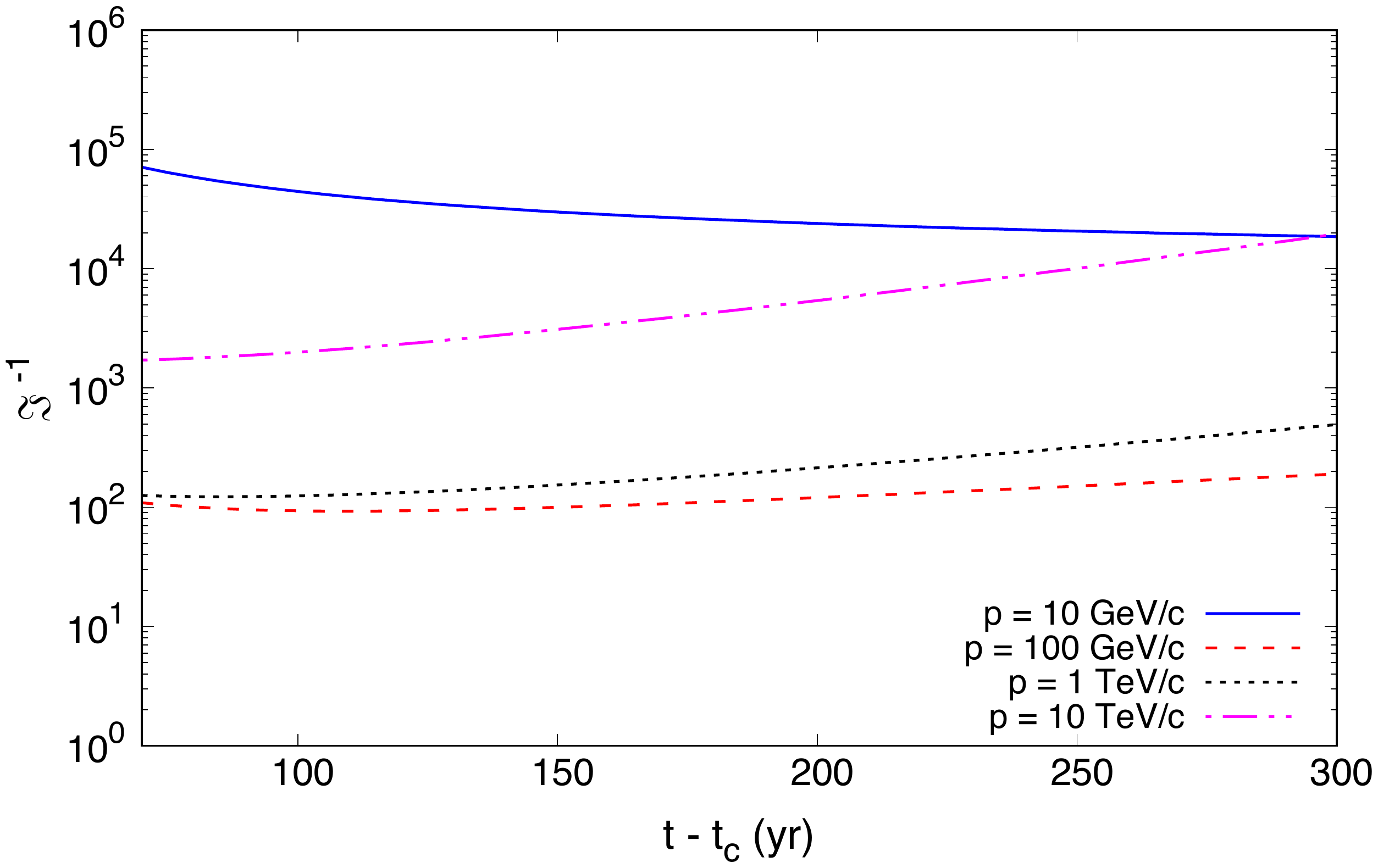}}
\caption{Ratio between self-generated diffusion coefficient $D_{\rm self}$ and Bohm diffusion coefficient $D_{\rm Bohm}$, as a function of the clump age for different particle momenta (from Eq.~\eqref{eq:D}, it follows that $\mathcal{F}^{-1} = (D_{\rm self}/D_{\rm Bohm})$). $D_{\rm self}$ is obtained by imposing $1/\Gamma_{\rm CR} = \min(1/\Gamma_{D},\tau_{\rm age})$. \textit{Left:} results in the clump skin, where $\Gamma_{D}=\Gamma_{\rm NLD}$ is considered. \textit{Right:} results for the clump interior, where $\Gamma_{D}=\Gamma_{\rm IND}$ is considered.}
\end{figure*}

The dominant mechanism of wave damping in the clump magnetic skin is the NLD, because we assume the plasma to be completely ionized. Its damping rate can be expressed as (see \citealp{nll})
\begin{equation}
\label{eq:nll}
\Gamma_\textrm{D} (k) = \Gamma_\textrm{NLD} (k) = (2c_k)^{-3/2} k v_A \sqrt{\mathcal{F}(k)}
\end{equation}
where $c_k=3.6$. In stationary conditions (when the system age is not a limiting factor), the wave growth rate (due to streaming instability) equals the damping rate, as $\Gamma_\textrm{CR} = \Gamma_\textrm{D}$. Equating Eq.~(\ref{eq:stream}) to Eq.~(\ref{eq:nll}), the power in the resonant turbulent momentum results in
\begin{equation}
\label{eq:nllF}
\mathcal{F} = \left[ \frac{16}{3} \frac{\pi^2}{B_0^2} \left(p^4 v(p) \frac{\partial f}{\partial r} \right)_{p=p_\textrm{res}} r_L \right]^{2/3} 2c_k
\end{equation}
We recall that, in Eq.~(\ref{eq:nllF}), the CR density gradient is computed within the clump skin, along the radial dimension. We need to verify whether the stationarity assumption is correct or not. Once we insert Eq.~(\ref{eq:nllF}) into Eq.~(\ref{eq:nll}), setting $v_A$ in $B_0=10 \mu$G and a typical ion density for a clump of $n_i=10^{-4}n_c=10^{-1}$~ions~cm$^{-3}$, we obtain that stationary is not valid for CR momenta larger than $p \geq 1$~TeV/c, where the clump age constraints the damping mechanism. Therefore, in this case, the power in turbulence is computed equating the growth rate of the MHD waves, as reported in Eq.~(\ref{eq:stream}), to the inverse of the clump age. This gives
\begin{equation}
\label{eq:Fage}
\mathcal{F} =  \frac{16}{3} \pi^2 \frac{v_A}{B_0^2} \left[ p^4 v(p) \frac{\partial f}{\partial r} \right]_{p=p_\textrm{res}} \tau_\textrm{age}
\end{equation}
The result of this computation is shown in Fig.~\ref{fig:nll}. The turbulence is generated in the clump magnetic skin, such that CRs with momentum between $100$~GeV/c and $1$~TeV/c are closer to the Bohm diffusive regime.   

On the contrary, in the clump interior, where neutral particles are abundant, the most efficient damping mechanism is IND (see \citealp{in} and \citealp{lara}). Waves dissipate energy because of the viscosity produced in the charge exchange between ions and neutrals, such that previously neutrals start to oscillate with the waves. The frequency of ion-neutral collision is \citep{kulsrud,druryMax}
\begin{equation}
\nu_\textrm{c} = n_\textrm{n} \langle \sigma v \rangle = 8.4 \times 10^{-9} \left(\frac{n_\textrm{n}}{1 \textrm{cm}^{-3}} \right) \left(\frac{T_\textrm{c}}{10^4 \, \textrm{K}} \right)^{0.4}
\end{equation}
where an average over thermal velocities is considered. The rate of IND depends on the wave frequency regime, namely whether ions and neutrals are strongly coupled or not. Defining the wave pulsation $\omega_k=kv_A$ in a collision-free medium, then the study of the dispersion relation defines different regimes for ion-neutral coupling depending on the value of the ion-to-neutral density ratio defined in Eq.~(\ref{eq:epsilon}). These regimes are as follows:
\begin{itemize}
\item[I)]  If $\epsilon < 1/8$, there's a range of $\omega_k$ for which waves can not propagate, that is a range of $k$ for which $\omega_k$ is a purely imaginary number. This range is for
\begin{equation}
4\epsilon < \frac{\omega^2_k}{\nu_c^2} < \frac{1}{4}
\end{equation}
which, within our assumptions ($\epsilon=1.5 \times 10^{-3}$), equals to CR momenta in $15 \, \textrm{GeV/c}< p < 95 \, \textrm{GeV/c}$;
\item[II)]  In the intervals $\epsilon \ll 1$ and $ (\omega_k/\nu_c)^2 \ll 4\epsilon$, then
\begin{equation}
\label{eq:inLow}
\Gamma_\textrm{IND} (k) = -\frac{\omega_k^2}{2 \nu_c} = -\frac{k^2 v_A^2}{2 \nu_c}
\end{equation}
\item[III)]  If $\epsilon \ll 1$ and $(\omega_k/\nu_c)^2 \ll 1/4$, then 
\begin{equation}
\label{eq:inHigh}
\Gamma_\textrm{IND} = - \frac{\nu_c}{2} 
\end{equation}
\item[IV)]  If $\epsilon \gg 1$, then 
\begin{equation}
\Gamma_\textrm{IND} (k) = - \frac{\nu_c}{2} \left[ \frac{(\omega^2_k/\nu_c^2)}{(\omega^2_k/\nu_c^2)+\epsilon^2} \right]
\end{equation}
\end{itemize}
\noindent
Again, the damping time $\Gamma_\textrm{D}^{-1}$ should be compared with  the clump age $\tau_\textrm{age}$. For $p=10$~GeV/c the IND time is shorter than the clump age, therefore setting the equilibrium condition $\Gamma_\textrm{D} = \Gamma_\textrm{IN}$ through Eq.~(\ref{eq:inHigh}), we find
\begin{equation}
\label{eq:inHighF}
\mathcal{F} = \frac{16}{3} \pi^2 \frac{v_A}{B_0^2} \frac{2}{\nu_c} \left[p^4 v(p) \frac{\partial f}{\partial r} \right]_{p=p_\textrm{res}} 
\end{equation}
For $p=100$~GeV/c, the dominant damping mechanism is still IND. Setting the equilibrium condition $\Gamma_\textrm{D} = \Gamma_\textrm{IN}$ through Eq.~(\ref{eq:inLow}), we get
\begin{equation}
\label{eq:inHighF}
\mathcal{F} = \frac{16}{3} \pi^2 \frac{1}{B_0^2} \left[p^4 v(p) \frac{\partial f}{\partial r} \right]_{p=p_\textrm{res}} \frac{2\nu_c}{k^2_\textrm{res} v_A} 
\end{equation}
On the other hand, for $p \geq 1$~TeV/c, the clump age is the limiting factor since IN damping requires a longer time. 
As shown in Fig.~\ref{fig:ind}, IND is mostly effective in damping waves resonant with CR particles of momentum lower than $10$~GeV/c. Nonetheless, a strong suppression of the diffusion coefficient is reached between $100$~GeV/c and $1$~TeV/c. 

\subsection{Perpendicular diffusion}
\label{subsec:perpD}
As shown in Figure~\ref{fig:MField6Obl}, the large scale magnetic field is compressed and stretched around a large fraction of the clump surface. In this region the Alfv\'enic turbulence produced by CR-driven instabilities is not enough to reach the Bohm limit because we have $\delta B_s \simeq (0.1-0.3) B_s$ at most, as shown in Fig.~\ref{fig:nll}. As a consequence, if additional pre-existing turbulence is not amplified at the same level of the regular field, the penetration inside the clump requires a perpendicular diffusion.
According to quasi linear theory, the diffusion perpendicular to the large scale magnetic field, $D_\perp$, is related to parallel diffusion, $D_\parallel$, through \citep{casse}
\begin{equation}
\label{eq:dperp1}
D_\perp = D_\parallel \frac{1}{1+(\lambda_\parallel/r_L)^2}
\end{equation}
where $\lambda_\parallel$ is the particle mean free path along the background field $B_0$. Since $\lambda_\parallel=r_L (\delta B/B_0)^{-2}$, the perpendicular diffusion coefficient results in
\begin{equation}
\label{eq:dperp}
D_\perp = D_\parallel \frac{1}{1+(\delta B / B_0)^{-4}}
\end{equation}
Hence, the radial diffusion into the clump is strongly suppressed with respect to the azimuthal diffusion along the field lines provided that a tiny amplification of the magnetic field is realized. In the following, we assume that in the region downstream of the shock $(\delta B(k) / B)_\textrm{down} \simeq 1$ at all scales $k$ resonant with accelerated particles. Our results of MHD simulations show that the magnetic field in the clump skin is amplified, reaching $\sim$10 times the value in the unperturbed downstream, so that $B_s=10B_\textrm{down}$ (Fig.~\ref{fig:MField6Obl}). If the downstream pre-existing turbulence is amplified in the skin as well, then the Bohm diffusion limit can be reached: this is the case for isotropic diffusion, where the distinction among parallel and transverse diffusion is lost ($D_\perp = D_\parallel$) and a strong suppression of the diffusion coefficient is realized. However, if the turbulence in the clump skin remains at the same level than the unperturbed downstream, then $(\delta B / B)_s \simeq 0.1$: this implies that in the skin parallel diffusion holds with $D_{\parallel s}=10 D_{\parallel \textrm{down}}$, while for the perpendicular diffusion, using Eq.(\ref{eq:dperp}), we get $D_{\perp s}=10^{-3} D_{\parallel \textrm{down}}$. Therefore, in this regime, particle penetration inside the clump is even more suppressed than in the case of isotropic diffusion.

\subsection{Proton spectrum}
\label{subsec:resultsP}
\begin{figure}
\centering
\includegraphics[scale=0.33]{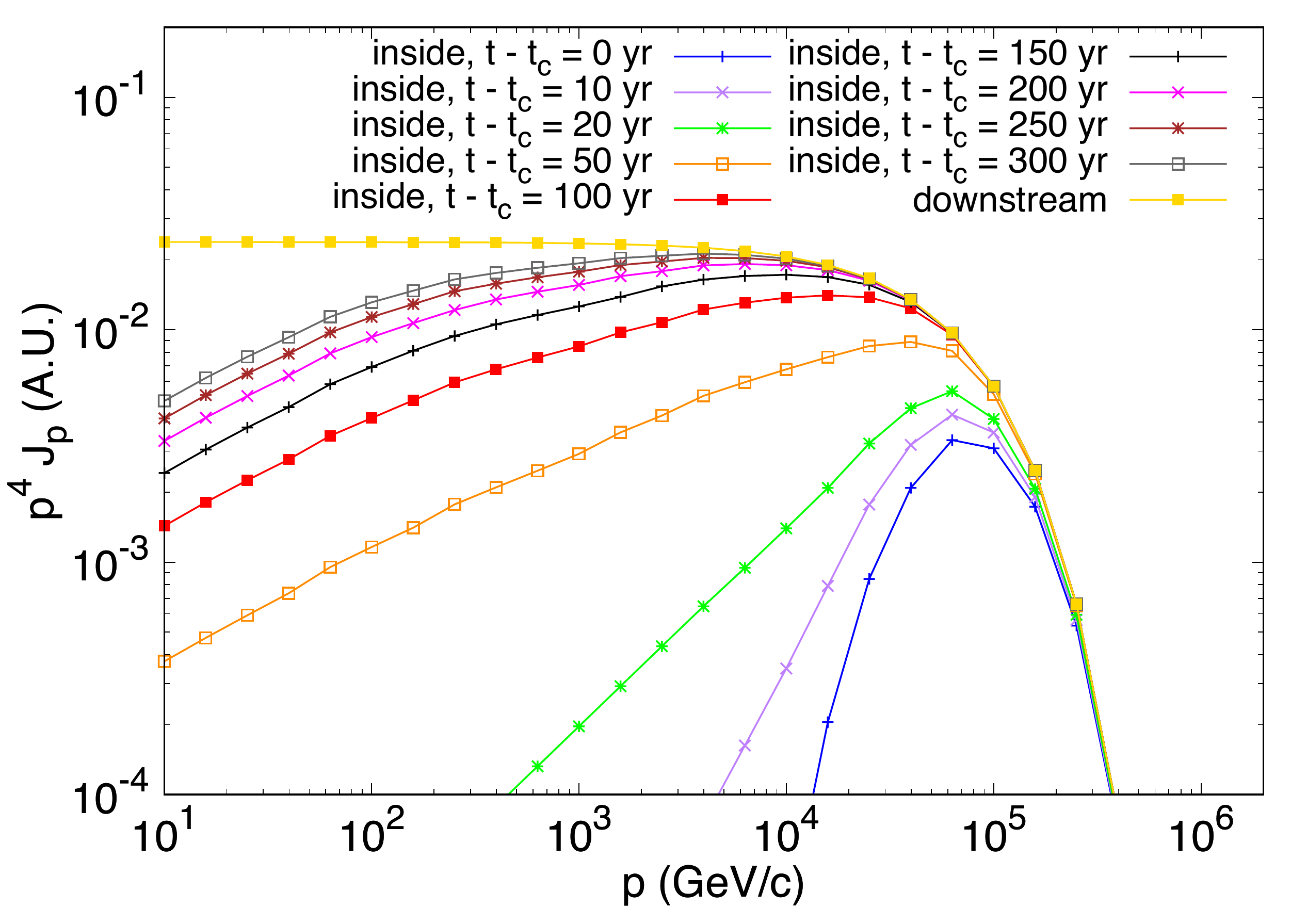}
\caption{Proton spectrum inside clumps of different ages. The particle density in the downstream, which is constant in time, is also reported.}
\label{fig:psIn} 
\end{figure}

Once the proton distribution function is known from the solution of Eq.~(\ref{eq:transport}), it is possible to obtain the proton energy spectrum inside the clump $J_c (p,t)$ at different times with respect to the first shock-clump contact, that we will indicate as $t=t_\textrm{c}$. The average spectrum inside the clump reads as
\begin{equation}
\label{eq:sp}
J_c (p,t)= \frac{1}{V_c} \frac{d^3N_c(t)}{dp^3}
\end{equation}
where $V_c=4\pi R_c^3/3$ is the clump volume and $d^3N_c(t)/dp^3$ is the number of protons inside the clump at a time $t$ per unit volume in momentum space. The spectrum can be computed summing upon all the discretized bins which define the clump volume. In this way, we obtain
\begin{equation}
\label{eq:spectrum}
J_c (p,t) =  \frac{2 \pi}{V_c} \sum_{i \in clump} f (r_i,z_i,p,t) r_i \Delta r_i \Delta z_i
\end{equation}
Results are shown in Fig.~\ref{fig:psIn}, where a proton cut-off momentum of $p_\textrm{cut}=70$~TeV/c was set in order to reproduce the very-high-energy gamma-ray data. Different proton spectra are expected at different times, namely for younger clumps the particle spectrum is much harder than the one accelerated at the shock as defined in Eq.~(\ref{eq:dsa}). This is explained by the prevention of penetration of low-energy CRs into the clump due to the amplified magnetic field at the skin and because of the linear dependency of the diffusion coefficient with the particle momentum. In this way, the entrance of these particles into the clump is delayed. The spectral index of protons below 100 GeV/c is as hard as $\alpha=-3.50$ when the clump age is $50$~yr, moving to $\alpha=-3.54$ when the clump age is $150$~yr, and finally $\alpha=-3.57$ when the clump is $300$~yr old. On the other hand, CRs with $p \gtrsim 100$~TeV/c are quite unaffected by the presence of the clump.

\section{Gamma rays from a uniform clump distribution inside the shell}
\label{sec:gamma}
If the number of clumps is large enough, they could be the main source of gamma-ray emission, due to hadronic inelastic collisions of CRs with the ambient matter. In such a case the gamma-ray spectrum would reflect the spectrum of particles inside the clumps rather than the one outside them, as produced by the shock acceleration. The effect of a clumpy environment on the time-dependent spectrum of gamma rays from a SNR has also been investigated by \citet{gaggero}. In this section we calculate the total gamma-ray spectrum due hadronic interactions, assuming that clumps are uniformly distributed over the CSM where the shock expands. For the clump density we assume here, the effect of a SNR shock impacting on a clump distribution can be described, with good accuracy, as the result of individual shock-clump interactions. Note that the average distance among clumps is much larger than the clump size and the simulation box.

The emissivity rate of gamma rays from a single clump, given the differential flux of protons inside the clump $\phi_c(T_p,t)$ and the density of target material, is
\begin{equation}
\label{eq:gammaF}
\epsilon_c(E_{\gamma},t)= 4\pi n_\textrm{c} \int dT_p \frac{d\sigma_{pp}}{dE_{\gamma}}(T_p,E_{\gamma}) \phi_c (T_p,t)
\end{equation}
where $d\sigma_{pp}/dE_{\gamma}$ is the differential cross section of the interaction while $\phi_c(T_p,t)$ is obtained from the spectrum in Eq.~(\ref{eq:spectrum}), $T_p$ being the particle kinetic energy. 
We used the analytical parametrization for the p-p cross section provided by the \text{LibPPgam} library (see \citealp{libpg}) and, specifically, we chose the parametrization resulting from the fit to \text{Sibyll 2.1}.

In order to evaluate the cumulative distribution resulting from a fixed distribution of clumps we need to include clumps that satisfy the following two conditions: I) they should survive (not evaporated); II) they should be located between the position of the contact discontinuity (CD) $R_{\textrm{cd}}$ and the shock position $R_s$. Indeed we will assume that, once a clump passes through the CD, either it is destroyed by MHD instabilities or it soon gets emptied of CRs. Therefore we should consider the minimum time between the evaporation time $\tau_{\textrm{ev}}$, and the time elapsed between the moment the clump crosses the forward shock and the moment it crosses the contact discontinuity $\tau_{\textrm{cd}}$. As estimated in \S~\ref{sec:clump}, the evaporation time is of the order of few times the cloud crossing time (see Eq.~(\ref{eq:taucc})). For the parameters we chose, this time is always larger than the SNR age. In the following, we will consider the conservative value of $\tau_{\textrm{ev}} = \tau_{\textrm{cc}}$. The CD radial position, instead, can be estimated imposing that all the compressed matter is contained in a shell of size $\Delta R = R_{\textrm{SNR}} - R_{\textrm{cd}}$, so that
\begin{equation}
\frac{4}{3} \pi R^3_{\textrm{SNR}} n_\textrm{up} = 4 \pi R^2_{\textrm{SNR}} \Delta R n_\textrm{down}
\end{equation}
which for strong shock amounts to $\Delta R = \frac{1}{12}R_{\textrm{SNR}}$. Therefore, the time that a clump takes to be completely engulfed in the CD is 
\begin{equation}
\label{eq:tcd}
\tau_{\textrm{cd}} = \frac{(2R_c+\Delta R)}{\frac{3}{4}v_s} 
\end{equation}
The oldest clumps in the remnant shell will therefore have an age
\begin{equation}
T_{\textrm{c,max}}=\textrm{min}(\tau_{\textrm{ev}},\tau_{\textrm{cd}})
\end{equation}
In the following, we will consider a uniform spatial distribution of clumps, with number density $n_0=0.2$~clumps pc$^{-3}$, inside the remnant of age $T_{\textrm{SNR}}$. Therefore, the total number of clumps at a distance among $r$ and $r+dr$ from the source is equal to 
\begin{equation}
\frac{dn(r)}{dr}=4 \pi n_0 r^2  \implies \frac{dn(t)}{dt}= 4 \pi n_0 r^2(t) v_s(t) 
\end{equation}
Furthermore, we will assume constant shock speed. We are interested in the number of clumps with a given age $t_{\textrm{age}}(r)=T_{\textrm{SNR}}-t_c(r)$. The number of clumps with an age between $t_{\textrm{age}}-\Delta t$ and $t_{\textrm{age}}$, namely $N(t_{\textrm{age}})$, is equal to the number of clumps that the shock has encountered between $T_{\textrm{SNR}}-t_{\textrm{age}}$ and $T_{\textrm{SNR}}-t_{\textrm{age}}+\Delta t$. It is equal to
\begin{equation}
\label{eq:Ntc}
N(t_{\textrm{age}})=4\pi n_0 \int_{T_{\textrm{SNR}}-t_{\textrm{age}}}^{T_{\textrm{SNR}}-t_{\textrm{age}}+\Delta t} r(t^\prime)^2 v_s(t^\prime) dt^\prime
\end{equation}
The total number of clumps with $t_{\textrm{age}} \leq T_{\textrm{c,max}}$ is equal to $N_c \simeq 440$, which corresponds to the total mass in clumps inside the remnant shell equal to $M_c \simeq 45 M_{\odot}$. Consequently the total gamma-ray emissivity due to these clumps is
\begin{equation}
\label{eq:gammaC}
\epsilon_c (E_\gamma,T_{\textrm{SNR}})= \sum_{t_\textrm{age}=0}^{T_{\textrm{c,max}}}  N (t_{\textrm{age}}) \epsilon_\gamma (E_\gamma,t_{\textrm{age}}) 
\end{equation}
We also account for the emissivity from the downstream region of the remnant $\epsilon_\textrm{down} (E_\gamma)$, which is constant with time. The gas target in the downstream is considered with an average density $\langle n_\textrm{down} \rangle$, that satisfies mass conservation in the whole remnant
\begin{equation}
\frac{4}{3} \pi R^3_{\textrm{SNR}} n_\textrm{up} = \frac{4}{3} \pi (R^3_{\textrm{SNR}} - R^3_{\textrm{cd}}) \langle n_\textrm{down} \rangle 
\end{equation}
Therefore we compute the gamma-ray flux from the source located at a distance $d$ as
\begin{equation}
\label{eq:gammaF}
\phi_\gamma (E_\gamma,T_\textrm{SNR})= \frac{1}{d^2} \left[ V_\textrm{c} \epsilon_c (E_\gamma,T_\textrm{SNR}) + V_\textrm{down} \epsilon_\textrm{down} (E_\gamma) \right]
\end{equation}
where $V_\textrm{down}=V_\textrm{shell}-N_\textrm{c} V_\textrm{c}$ and $V_\textrm{shell}= 4 \pi (R^3_{\textrm{SNR}} - R^3_{\textrm{cd}})/3$.

\section{Application to RX J1713.7-3946}
\label{sec:1713}
The Galactic supernova remnant RX J1713.7-3946 (also called G347.3-0.5) represents one of the brightest TeV emitters on the sky. The origin of its gamma-ray flux in the GeV-TeV domain (see \citealp{fermi} and \citealp{hess}) has been the object of a long debate, since both hadronic and leptonic scenarios are able to reproduce, under certain circumstances, the observed spectral hardening. The presence of accelerated leptons is guaranteed by the detected X-ray shell (see \citealp{slane} and \citealp{tanaka}), which shows a remarkable correlation with the TeV gamma-ray data, indicating a strong link between the physical processes responsible for these emission components. Meanwhile, a clear signature of accelerated hadrons, that would come from neutrinos, is still missing. RX J1713.7-3946 has been used as a standard candidate for the search of a neutrino signal from Galactic sources \citep{kappes, vissaniRxj, morlinoRxj}. One of the arguments against the hadronic origin of the radiation has been claimed because of the absence of thermal X-ray lines (see \citealp{katz} and \citealp{ellison}). In the scenario where the remnant is expanding into a clumpy medium, the non observation of a thermal X-ray emission is naturally explained by the low density plasma between clumps. On the other hand, since clumps remain mainly unshocked and therefore cold, they would not be able to emit thermal X-rays. 

Clearly, the distribution of gas in RX J1713.7-3946 is crucial to establish the origin of the observed gamma rays. The target material required by pp interactions may be present in any chemical form, including both the molecular and the atomic gas. High-resolution mm-wave observations of the interstellar CO molecules with NANTEN \citep{fukui} revealed the presence of molecular clouds in spatial correlations with TeV gamma rays, in the northwestern rim of the shell. The densest cores of such clouds have been detected in highly excited states of the molecular gas, manifesting signs of active star formation, including bipolar outflow and possibly embedded infrared sources \citep{sano}. Other density tracers, such as Cesium, have also confirmed the presence of very dense gas in the region ($n > 10^4$~cm$^{-3}$), as indicated in a recent MOPRA survey \citep{mopra}. Furthermore, a combined analysis of CO and HI \citep{fukui12} shows a counterpart of the southeastern rim of the gamma-ray data in atomic hydrogen. The multi-wavelength observations point towards the clear presence of a non homogeneous environment, where the young SNR is expanding. 

The estimated distance of the remnant is about $d \simeq 1$~kpc (\citealp{fukui,moriguchi2005}), while the radial size of the detected gamma-ray shell today extends up to $R_s \simeq 0.6$~deg (\citealp{hess}). The remnant is supposed to be associated to the Chinese detected type II SN explosion of 393 AD (\citealp{china}); this would assign to the remnant an age of $T_\textrm{SNR} \simeq 1625$~yr. The age, distance and detected size yield an average shock speed of about $\langle v_s \rangle \simeq 6.3 \times 10^8$~cm~s$^{-1}$. Measurements of proper motion of X-ray structures indicate that the shock speed today should be $v_s \leq 4.5 \times 10^8$~cm/s (\citealp{uchiyama}), meaning that the shock has slightly slown down during its expansion. This is expected in SNR evolution \citep{truelove} during both the ejecta-dominated (ED) and the Sedov-Taylor (ST) phases. RX J1713.7-3946 is nowadays moving from the ED to the ST phase, therefore we can safely assume a constant shock speed through the time evolution up to now, with a value of $v_s = 4.4 \times 10^8$~cm~s$^{-1}$ (\citealp{stefanofelix}). At this speed, the time that the CD takes to completely engulf a clump is, following Eq.~(\ref{eq:tcd}), $\tau_{\textrm{cd}}  \simeq 300 \, \textrm{yr}$. On the other hand, the evaporation time would be much longer, indicating that the relevant clumps contributing to the gamma-ray emission are younger than $T_{\textrm{c,max}}=300$~yr. 

\begin{figure}
\centering
\includegraphics[scale=0.45]{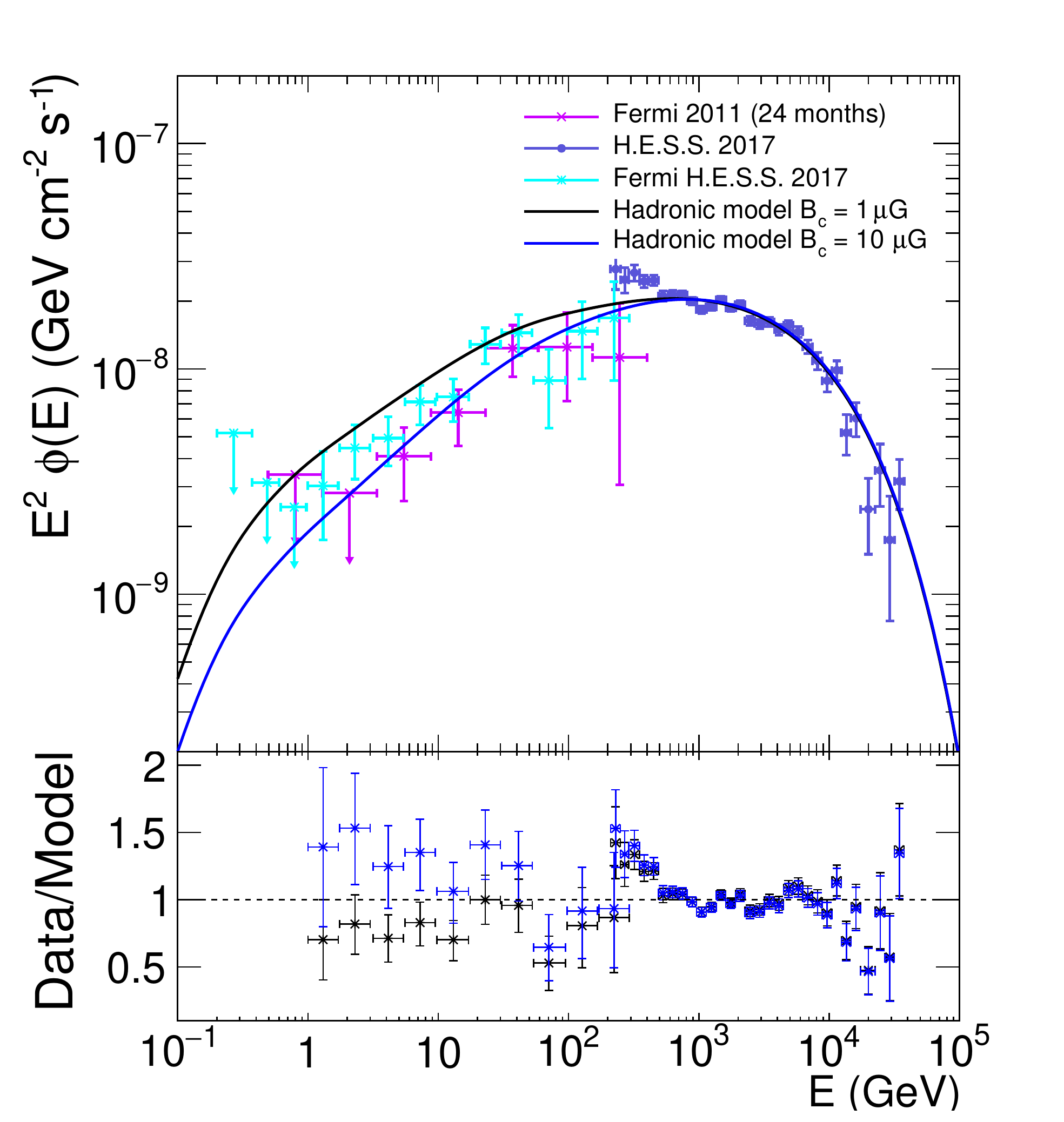}
\caption{Gamma-ray flux from SNR RX J1713.7-3946. The points are Fermi-LAT data (pink), H.E.S.S. data (violet) and H.E.S.S. analysis of Fermi-LAT data (light blue). The hadronic models (solid lines) refer to the configuration with a magnetic field inside the clump equal to $B_c = 1 \, \mu$G (black) and to $B_c = 10 \, \mu$G (blue). The field in the clump skin is fixed to $B_s = 100 \, \mu$G in both models.}
\label{fig:models} 
\end{figure}

With the parameters representing RX J1713.7-3946, as defined above, we can compute the gamma-ray flux of the remnant shell, through Eq.~(\ref{eq:gammaF}). We fix the normalization $k$ of the resulting gamma-ray flux by minimizing the $\chi^2$ of the Fermi-LAT and H.E.S.S. data with respect to our model. We investigate two different configurations. The first model explores a configuration with the magnetic field inside the clump reduced by a factor of $10$ with respect to the CSM value, in order to account for the effect of IND, therefore it is set to $B_c = 1 \, \mu$G. The second model, instead, explores the situation where no IND is acting, therefore the magnetic field inside the clump is set to be $B_c = 10 \, \mu$G, as in the CSM. Results are shown in Fig.~\ref{fig:models}. The GeV data from two years of data-taking of the Fermi-LAT satellite \citep{fermi} are reported, together with the H.E.S.S. TeV data \citep{hess} and with the H.E.S.S. Collaboration analysis of five years of Fermi-LAT data, as reported in \citet{hess}. The two models predict slightly different trend in the GeV emission of the remnant. A more pronounced hardening in the case of $B_c = 10 \, \mu$G better reproduces the GeV data, while a flatter trend is visible in case diffusion would act less efficiently inside the clump. In this respect, electrons are more suitable to derive constraints on the magnetic field properties of the remnant. A more quantitative study on secondary electrons from $pp$ interactions will be discussed elsewhere. 

The normalization constant $k$, obtained by fitting the gamma-ray data, defines the amount of ram pressure $P_\textrm{ram}=\rho_\textrm{up} v_s^2$ that is instantaneously converted into CR pressure. The latter is defined, for relativistic particles, as
\begin{equation}
  P_{\textrm{CR}} = \frac{k}{3} \int_{m_pc}^{\infty} 4 \pi p^2 dp f_0(p) pc \,.
\end{equation}
The efficiency of the pressure conversion mechanism from bulk motion to accelerated particles equals to $\eta = P_{\textrm{CR}}/P_\textrm{ram} \simeq 2$\%. Such a value is somewhat smaller than the efficiency estimated by other works in the context of hadronic scenarios, where usually $\eta \simeq 10-20\%$ \cite[see, e.g][]{Morlino2009, stefanofelix}. Compared to \cite{Morlino2009} the main differences are due to the highest total target mass we use here ($\sim 45 M_{\odot}$ vs. $\sim 15 M_{\odot}$) which is close to the total mass in clumps estimated by HD simulations \citep{inoue}.
A weaker effect is also due to the fact that we are neglecting adiabatic losses, which leads to a smaller acceleration efficiency by less than a factor of two.
Also, comparing our result with \citet{stefanofelix} we should note a few more differences. We consider a constant shock speed and assume a $\propto p^{-4}$ accelerated spectrum while \citet{stefanofelix} use a time dependent shock velocity and a steeper acceleration spectrum $\propto p^{-4.2}$. The latter assumption implies a number density of accelerated  protons at 100 TeV smaller by a factor of $\sim 10$ (for the given acceleration efficiency). For this reason, \citet{stefanofelix} adopted a larger target mass in clumps, $\sim 500 M_{\odot}$. 

\subsection{Observing clumps through molecular lines}
\label{subsec:mlines}
An interesting possibility to detect clumps is provided by radio observations. Secondary electrons emit synchrotron radiation in the radio domain. On top of this continuum, the molecular gas emits lines. For instance, rotational CO lines are often observed in these systems \citep{fukui}. In the case of RX J1713.7-3946, a few arcsecond angular resolution is needed to probe the spatial scales of clumps: such a small scale can currently be achieved only through the superior angular resolution of the Atacama Large Millimeter Array (ALMA). A precise pointing is however required, since the instrument field of view of $\leq 35''$ would not entirely cover a region as extended as the remnant RX J1713.7-3946. 

In the following, we will evaluate the radio flux for the $J=1\rightarrow 0$ rotational line of the CO molecule. This transition is located at $\nu=115$~GHz (band 3 of ALMA receivers) and radiates photons with a rate equal to $A_{10}=6.78 \times 10^{-8}$~Hz. Assuming a CO abundance of $n_\textrm{CO}/n_c=7 \times 10^{-5}$ and a clump density of $n_c=10^3$~cm$^{-3}$, the expected flux from an individual clump amounts to
\begin{equation}
\label{eq:fRadio}
F=\frac{h A_{10}}{4\pi d^2} N_\textrm{CO} = 3.15 \times 10^{-3} \, \textrm{Jy}
\end{equation}
where $h$ is the Planck constant and $N_\textrm{CO}$ is the number of CO molecules contained in each clump ($d=1$~kpc is assumed). The flux level obtained in Eq.~(\ref{eq:fRadio}) is well within the performances of the nominal ten $7-$m diameter antennas configuration of the ALMA observatory. In this respect, CO line measurements constitute a powerful tool to constrain the density of individual clumps located in the remnant environment.

Finally, a well known shock tracer into molecular clouds is constituted by the SiO molecule. Si ions are generally contained in the dust grains, which are destroyed by the passage of a shock, and can form SiO in gas phase \citep{Gusdorf2008A, Gusdorf2008B}. This tracer has been successfully used for the supernova remnant W~51C \citep{Dumas2014}. In the case of  RX J1713.7-3946, existing observations toward the north-west rim, in the so-called Core C, do not show evidence for significant amount of SiO emission \citep{mopra}. This result can be interpreted in different ways: either the Core C is located outside of the remnant or, if it is inside, the shock is not yet penetrated into the densest cores where the dust grains are typically located. A further possibility is that the shock propagating inside the core is not strong enough to efficiently sputter Si ions from the grains. Indeed \citet{Gusdorf2008A} showed that a shock velocity $\gtrsim 25$ km s$^{-1}$ is necessary for an efficient sputtering and  the shock speed inside the clump could be lower than such a threshold if $\chi \gtrsim 10^5$ (see Eq.~(\ref{eq:vsc})). As a consequence, a positive SiO detection could shed light on the value of the density contrast $\chi$.

\subsection{Clumps and the X-ray variability}
\label{subsec:xray}
The origin of bright hot spots in the non-thermal X-ray image of RX J1713.7-3946 \citep{uchiyama}, decaying on the time scale of one year, remains an unexplained puzzle. Several time-variable compact features have been identified in the Chandra data, mostly in the northwest part of the shell. The size of the fast variable X-ray hot spots in RX J1713.7-3946 is about $\theta \simeq 20$~arcsec, which corresponds to a linear size of $L_x = d \theta \simeq 3 \times 10^{17}$~cm at a distance  $d \simeq 1$~kpc. Similar time-dependent features have also been identified in the young SNRs Cas~A \citep{casa,sato18,fraschetti18} and G~330.2+1.0 \citep{borowsky}. It has been already suggested that the X-ray variability could be connected to the clump scenario applied at the forward shock \citep{inoue} or, for the case of Cas~A, even at the inward shocks originated as a reflection from the collision among the forward shock and overdense clumps \citep{fraschetti18}. The striking similarity between the physical size of the observed hot spots and the MHD instabilities which are formed in shock flows around clumpy structures suggests a possible intrinsic physical link.

In the context of shock propagation into a non uniform ambient medium, as discussed above, a natural interpretation of the X-ray variability seems the electron synchrotron cooling in the amplified large scale magnetic field. It was shown in Fig.~\ref{fig:MField6Obl}, that the spatial scale where amplification takes place is of the order of the clump size. The time scale over which electrons of energy $E$ lose energy is $t_\textrm{synch} \simeq 12.5 (B/\rm{mG})^{-2} (E/\rm{TeV})^{-1}$~yr. The typical energy of synchrotron photons is $E_{\rm synch} = 0.04  (B/\rm{mG}) (E/\rm{TeV})^2$~keV, hence the energy loss timescale at the observed frequency is
\begin{equation}
t_{\rm synch} \simeq 2.4 \left( \frac{B}{\rm{mG}} \right)^{-3/2} \left(\frac{E_{\rm syn}}{\rm{keV}} \right)^{-1/2}  {\rm yr} \, .
\end{equation}
For a density contrast of $\chi=10^5$ (see Eq.~(\ref{eq:chi})), our MHD simulation shows that the amplification of magnetic field can bring the background field up to 10 times above the downstream value. Hence a mG magnetic field needed to explain the observed timescale would require a magnetic field in the downstream of $\sim 100 \mu$G.

\subsection{Resolving the gamma-ray emission}
\label{subsec:CTA}
\begin{figure}
\centering
\includegraphics[scale=0.42]{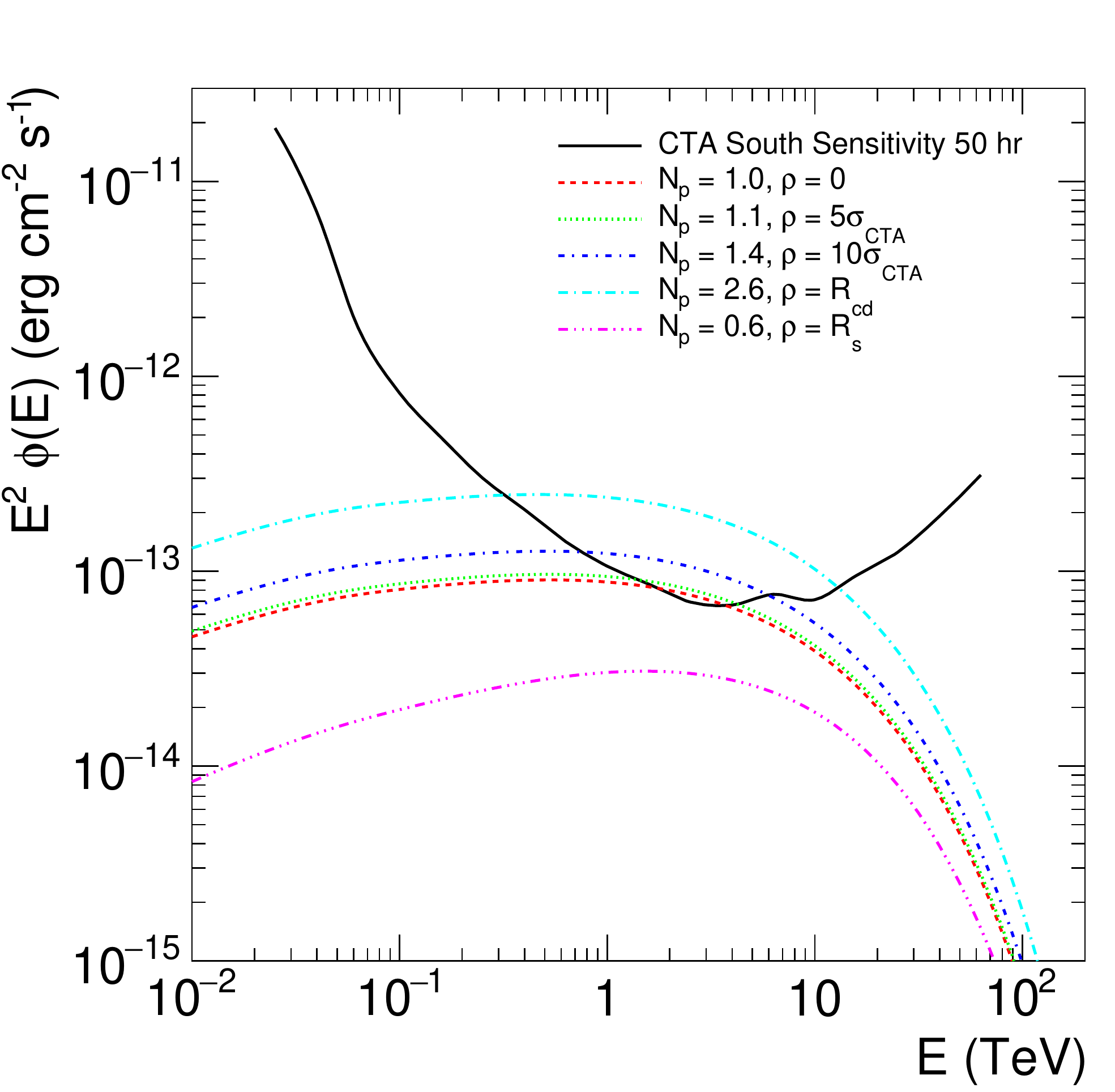}
\caption{CTA Southern array sensitivity curve for point-like sources located in the FoV center (zenith $\theta=20$~deg, pointing average) for an observation time equal to 50 hr (black solid line). Also shown are the gamma-ray fluxes due to the overlapping clumps (indicated in the legend) in different circular sky regions of radius $\sigma_\textrm{CTA}$, located at a distance $\rho$ from the SNR center.}
\label{fig:clumpCTA} 
\end{figure}

The deep morphological and spectroscopic studies of SNRs are among the highest priority scientific goals of the forthcoming Cherenkov telescope Array (CTA) \citep{cta,ctaRXJ}. Despite its great potential, CTA will be not able to resolve the gamma-ray emission from individual clumps. For a SNR at a distance of $1$~kpc, the angular extension of a clump with a typical size of $\sim 0.1$~pc does not exceed $20$~arcsec, which is one order of magnitude smaller than the angular resolution of CTA.  Nevertheless, the component related to the superposition of gamma-ray emission of several clumps within the instrument point spread function (PSF) in principle could be detected. 
Hence, we estimate the number of clumps $dN_\textrm{p}/d\rho$ which overlap along the line of sight $l$, when observing the disc-like region with radius equal to the instrument PSF centered at a distance $\rho=\sqrt{r^2-l^2}$ from the center of the SNR. Assuming a uniform distribution of clumps, as described in \S~\ref{sec:gamma}, and integrating it along the line of sight, we obtain
\begin{equation}
\label{eq:np}
\frac{dN_\textrm{p}}{d\rho} (\rho) = 2 n_0 \left( \sqrt{R^2_\textrm{s}-\rho^2}-\max[0,\sqrt{R^2_\textrm{cd}-\rho^2}] \right)
\end{equation}
where the emission is assumed to come from the shocked ISM located between the contact discontinuity and the forward shock \citep{Morlino2012}. Considering a uniform map of the whole remnant by CTA, we compute the gamma-ray flux from several circular regions centered at a given $\rho$ and with radius equal to $\sigma_\textrm{CTA}=0.037$~deg \footnote{\url{http://www.cta-observatory.org/science/cta-performance/}} (at $E_\gamma = 10$~TeV) \citep{ctaScience}. We considered $\rho$ spanning from $0$ to $R_s$. Moreover, we should take into account the different ages of clumps, since they produce gamma rays with different spectral shapes, as shown in \S~\ref{subsec:resultsP}. These fluxes are represented in Fig.~\ref{fig:clumpCTA}, where also shown is the sensitivity curve of the CTA Southern array, for a 50 hr observation of a point-like source centered in the instrument field of view (FoV). The predicted flux clearly shows that CTA will be able to resolve the gamma-ray emission from clumps contained in a circle of radius equal to its high-energy PSF over about one decade in energy. However, the gamma-ray fluxes expected at different pointing regions strongly reflect the number of overlapping clumps, which represent the main contributors to the emission. Such a number is maximum in correspondence of $\rho=R_\textrm{cd}$, where $N_\textrm{p}=2.6$. Given the limited number of overlapping clumps in each pointing region, large fluctuations are expected, according to the Poissonian statistics. Therefore, the detection of such fluctuations constitutes a characteristic signature of the presence of clumps. The amount of the fluctuations depends on the clump density $n_0$ in the CSM. In fact, once the mass of the target gas is fixed, more massive clumps with $n_c=10^4$~cm$^{-3}$ would require a lower clump density and therefore would produce much stronger fluctuations on the scale of $\sigma_{\rm CTA}$.
Such kind of morphological studies are hence crucial to derive constraints on the number density of clumps in the remnant region. 
Large fluctuations on the scale of $\sigma_{\rm CTA}$ are not expected if the SNR is expanding into a uniform medium or into a medium where the density contrast is such that the clump evaporates soon after the shock crossing, namely if $\tau_{\rm ev} \ll T_{\rm SNR}$. The latter condition can be rearranged using Eqs.~(\ref{eq:vsc}) and (\ref{eq:taucc}) to give  an upper limit for the density contrast which reads $\chi \ll  (T_{\rm SNR} v_s/ 2 R_c)^2 \simeq 10^3 (R_c / 0.1 \, \textrm{pc})^{-2}$. Such a small density contrast also implies a smaller amplified magnetic field and, as a consequence, a flattening of the gamma-ray spectrum. 

\section{Conclusions}
\label{sec:concl}
The presence of inhomogeneities in the CSM, in the form of dense molecular clumps where a shock propagates, strongly affects the plasma properties, in such a way that the large scale magnetic field around the clumps results amplified. As a consequence, the propagation of particles accelerated at the shock proceeds in such a way that low-energy particles need more time to penetrate the clump compared to high-energy ones. The resulting energy spectrum of particles inside the clump is significantly harder than the spectrum accelerated at the shock through DSA. 
Such a scenario is very common in the case of core-collapse SNe, where the remnant typically expands in a region populated by dense molecular clouds. It is then necessary to account for the inhomogeneous CSM to correctly predict the gamma-ray spectrum from the SNRs.
In this work we numerically solved the propagation of accelerated particles in the shock region at the presence of clumps, taking into account the magnetic filed amplification produced by both the field compression and the shear motion, and its effect on the particle diffusion. The most important parameter in this scenario is the density contrast between the diffuse CSM and the clumps. Using MHD simulation we showed that the magnetic amplification is effective only when the density contrast is larger than $10^3$.

Given that clumps contain most of the target gas, the gamma-ray spectrum produced in hadronic collisions of accelerated particles appears to be much harder than the parent spectrum. We demonstrate this effect for the brightest SNR in TeV gamma rays, RX J1713.7-3946. The cumulative contribution of clumps embedded between the contact discontinuity and the current shock position is able to reproduce the observed GeV hardening, though some degeneracy in the parameter space of the model is present. 
Remarkably, for the gas density inside the clump assumed here ($n_c= 10^3$~cm$^{-3}$) the evaporation time is much longer than the SNR age. As a consequence, the clumps crossed by the forward shock do not produce significant thermal X-ray emission, in agreement with observations. We argue that such a scenario can naturally account for the fast variability in non-thermal X-ray reported in some hot-spots inside RX J1713.7-3946, thanks to the magnetic field amplified around the clumps. The electrons entering these regions rapidly lose their energy because of synchrotron emission. An independent signature on the \textquoteleft clump\textquoteright $\,$ origin of the gamma-ray emission could be revealed from a morphological study performed with the next generation of ground-based gamma-ray observatory CTA. The superb sensitivity and the high angular resolution of CTA could allow to resolve regions small enough that contains only few clumps. This implies that we do expect a large spatial fluctuation of the gamma-ray flux, unlike a scenario where the SNR is expanding into a uniform medium.

\section*{Acknowledgement}
\vspace{-0.2cm}
We acknowledge the author of the PLUTO code \citep{pluto} that has been used in this work for the modeling of  the shock-clump MHD interaction. This research also made use of the CTA instrument response functions (prod3b-v1), as released by the CTA Consortium and Observatory. SG acknowledges support from Agence Nationale de la Recherche (grant ANR- 17-CE31-0014) and from the Observatory of Paris (Action F\'ed\'eratrice CTA).

\bibliographystyle{mnras}

\begin{thebibliography}{99}
\bibitem[\protect\citeauthoryear{Abdalla et al.}{2016}]{hess} Abdalla H. et al. [H.E.S.S. Collaboration], 2016, Astron. \& Astrophys., 612, 6
\bibitem[\protect\citeauthoryear{Abdo et al.}{2011}]{fermi} Abdo A.~A. et al. [Fermi LAT Collaboration], 2011, Astrophys.~J., 734, 28
\bibitem[\protect\citeauthoryear{Acero et al.}{2017}]{ctaRXJ} Acero F. et al. [CTA Consortium], 2017, Astrophys.~J., 840, 74
\bibitem[\protect\citeauthoryear{Acharya et al.}{2017}]{ctaScience} Acharya B.~S. et al. [CTA Consortium], 2017, \url{arXiv:1709.07997}
\bibitem[\protect\citeauthoryear{Ackermann et al.}{2013}]{Ackermann2013} Ackermann M. et al. [Fermi LAT Collaboration], 2013, Science, 339, 807
\bibitem[\protect\citeauthoryear{Actis et al.}{2011}]{cta} Actis M. et al. [CTA Consortium], 2011, Exp.~Astron., 32, 193
\bibitem[\protect\citeauthoryear{Berezhko et al.}{2013}]{berezko} Berezhko E.~G., Ksenofontov L.~T., V\"olk H.~J., 2013, Astrophys.~J., 763, 14B
\bibitem[\protect\citeauthoryear{Blanford \& Ostriker}{1978}]{blanford} Blandford R.~D., Ostriker J.~P., 1978, Astrophys.~J.~Lett., 221, L29
\bibitem[\protect\citeauthoryear{Bell}{1978}]{bell} Bell A.~R., 1978, Mon.~Not.~R.~Astron.~Soc.,182, 147
\bibitem[\protect\citeauthoryear{Borowsky}{2018}]{borowsky} Borkowski, K.~J., Reynolds, S.~P., Williams, B.~J., Petre, R., 2018, Astrophys.~J.~Lett., 868, L21
\bibitem[\protect\citeauthoryear{Cardillo, Amato \& Blasi}{2016}]{cardillo} Cardillo M., Amato E., Blasi P., 2016, Astron. \& Astrophys., 595A, 58C
\bibitem[\protect\citeauthoryear{Casse, Lemoine \& Pelletier}{2002}]{casse} Casse F., Lemoine M., Pelletier G., 2002, Phys.~Rev.~D, 65, 023002
\bibitem[\protect\citeauthoryear{Chevalier}{1999}]{chevalier} Chevalier R.~A., 1999, Astrophys.~J., 511, 798
\bibitem[\protect\citeauthoryear{Drury}{1983}]{drury} Drury L.~O'C., 1983, Rep.~Prog.~Phys., 46, 973
\bibitem[\protect\citeauthoryear{Drury, Duffy \& Kirk}{1996}]{druryMax} Drury L.~O'C., Duffy P., Kirk J.~G., 1996, Astron. \& Astrophys., 309, 1002
\bibitem[\protect\citeauthoryear{Dumas et al.}{2014}]{Dumas2014} Dumas G., Vaupr\'e S., Ceccarelli C., Hily-Blant P., Dubus G., Montmerle T., Gabici S., 2014, Astrophys.~J.~Lett., 786, L24
\bibitem[\protect\citeauthoryear{Dwarkadas}{2007}]{dwarkadas} Dwarkadas V.~V., 2007, Astrophys.~J., 667, 226
\bibitem[\protect\citeauthoryear{Ellison et al.}{2010}]{ellison} Ellison D.~C., Patnaude D.~J., Slane P., Raymond J., 2010, Astrophys.~J., 712, 287
\bibitem[\protect\citeauthoryear{Fraschetti}{2013}]{fraschetti} Fraschetti F., 2013, Astrophys.~J., 770, 84
\bibitem[\protect\citeauthoryear{Fraschetti}{2018}]{fraschetti18} Fraschetti F., Katsuda S., Sato T., Jokipii J. R., Giacalone J., 2018, Phys.~Rev.~Lett., 120, 251101
\bibitem[\protect\citeauthoryear{Fukui et al.}{2003}]{fukui} Fukui Y. et al., 2003, Pub.~Astron.~Soc.~J., 55, L61
\bibitem[\protect\citeauthoryear{Fukui et al.}{2012}]{fukui12} Fukui Y. et al., 2012, Astrophys.~J., 746, 82
\bibitem[\protect\citeauthoryear{Gabici \& Aharonian}{2014}]{stefanofelix} Gabici S., Aharonian F.~A., 2014, Mon.~Not.~R.~Astron.~Soc.~Lett., 455, L70
\bibitem[\protect\citeauthoryear{Gabici \& Aharonian}{2016}]{GabiciAharonian2016} Gabici S., Aharonian F.~A., 2016, EPJ Web of Conferences, 121, 04001
\bibitem[\protect\citeauthoryear{Gabici \& Montmerle}{2016}]{gabicimontmerle} Gabici S., Montmerle T., 2016, PoS ICRC2015, 029
\bibitem[\protect\citeauthoryear{Gaggero et al.}{2018}]{gaggero} Gaggero D., Zandanel F., Cristofari P., Gabici S., 2018, Mon.~Not.~R.~Astron.~Soc., 475, 5237G
\bibitem[\protect\citeauthoryear{Giacalone \& Jokipii}{2007}]{giacalone} Giacalone J., Jokipii J., 2007, Astrophys.~J., 663, L41
\bibitem[\protect\citeauthoryear{Ginzburg \& Syrovatsky}{1961}]{ginzburg} Ginzburg V.~L., Syrovatsky S.~I., 1961, Progr.~Theor.~Phys.~Suppl., 20, 1-83
\bibitem[\protect\citeauthoryear{Gusdorf et al.}{2008a}]{Gusdorf2008A} Gusdorf A., Cabrit S., Flower D.~R., Pineau Des For\^{e}ts G., 2008, Astron. \& Astrophys., 482, 809
\bibitem[\protect\citeauthoryear{Gusdorf et al.}{2008b}]{Gusdorf2008B} Gusdorf A., Pineau Des For\^{e}ts G., Cabrit S., Flower D.~R., 2008, Astron. \& Astrophys., 490, 695
\bibitem[\protect\citeauthoryear{Hennebelle}{2013}]{hennebelle} Hennebelle P., 2013, New Trends in Radio Astronomy in the ALMA Era, 476, 115
\bibitem[\protect\citeauthoryear{Inoue et al.}{2012}]{inoue} Inoue T., Yamazaki R., Inutsuka S., Fukui Y., 2012, Astrophys.~J., 744, 71 
\bibitem[\protect\citeauthoryear{Kafexhiu et al.}{2016}]{libpg} Kafexhiu E., Aharonian F.~A., Taylor A.~M., Vila G.~S., 2014, Phys.~Rev.~D, 90, 12
\bibitem[\protect\citeauthoryear{Kappes et al.}{2007}]{kappes} Kappes A., Hinton J., Stegmann C., Aharonian F.~A., 2007, Astrophys.~J., 656, 870
\bibitem[\protect\citeauthoryear{Katz \& Waxman}{2008}]{katz} Katz B., Waxman E., 2008, J.~Cosm.~Astropart.~Phys. 01, 018
\bibitem[\protect\citeauthoryear{Klein, McKee \& Colella}{1994}]{klein} Klein R.~I., McKee C.~F., Colella P., 1994, Astrophys.~J., 420, 213
\bibitem[\protect\citeauthoryear{Kulsrud \& Cesarsky}{1971}]{kulsrud} Kulsrud R.~M., Cesarsky C.~J., 1971, Astrophys.~Lett., 8, 189
\bibitem[\protect\citeauthoryear{Mac Low at al.}{1994}]{MacLow94} Mac Low M.-M., McKee C.~F., Klein R.~I., Stone J.~M., Norman M.~L., 1994, Astrophys.~J., 433, 757
\bibitem[\protect\citeauthoryear{Jones at al.}{1996}]{Jones96} Jones T.~W., Ryu D., Tregillis I.~L. 1996, Astrophys.~J., 473, 365
\bibitem[\protect\citeauthoryear{Maxted et al.}{2012}]{mopra} Maxted N.~I. et al., 2012, Mon.~Not.~R.~Astron.~Soc., 422, 2230
\bibitem[\protect\citeauthoryear{Mignone et al.}{2007}]{pluto} Mignone A. et al., 2007, Astrophys.~J.~Suppl.~Ser., 170, 1 
\bibitem[\protect\citeauthoryear{Moriguchi et al.}{2005}]{moriguchi2005} Moriguchi Y., Tamura K., Tawara Y., Sasago H., Yamaoka K., Onishi T., Fukui Y., 2005, Astrophys.~J., 631, 947
\bibitem[\protect\citeauthoryear{Morlino, Amato \& Blasi}{2009A}]{Morlino2009} Morlino G., Amato E., Blasi P., 2009, Mon.~Not.~R.~Astron.~Soc., 392, 240
\bibitem[\protect\citeauthoryear{Morlino, Blasi \& Amato}{2009B}]{morlinoRxj} Morlino G., Blasi P., Amato E., 2009, Astropart.~Phys., 31, 376
\bibitem[\protect\citeauthoryear{Morlino \& Caprioli}{2012}]{Morlino2012} Morlino G., Caprioli D., 2012, Astron. \& Astrophys., 538, A81, 15
\bibitem[\protect\citeauthoryear{Nava et al.}{2016}]{lara} Nava L., Gabici S., Marcowith A., Morlino G., Ptuskin V.~S., 2016, Mon.~Not.~R.~Astron.~Soc., 461, 4
\bibitem[\protect\citeauthoryear{Orlando et al.}{2005}]{orlando2005} Orlando S., Peres G., Reale F., Bocchino F., Rosner R., Plewa T., Siegel A., 2005, Astron. \& Astrophys., 444, 505
\bibitem[\protect\citeauthoryear{Orlando et al.}{2008}]{orlando2008} Orlando S., Bocchino F., Reale F., Peres G., Pagano P., 2008, Astrophys.~J., 678. 274
\bibitem[\protect\citeauthoryear{Padovani, Galli \& Glassgold}{2009}]{padovani} Padovani M., Galli D., Glassgold A.~E., 2009, Astron. \& Astrophys., 501, 2
\bibitem[\protect\citeauthoryear{Phan, Morlino \& Gabici}{2018}]{Phan2018} Phan V.~H.~M., Morlino G., Gabici S., 2018, Mon.~Not.~R.~Astron.~Soc., 480, 5167 
\bibitem[\protect\citeauthoryear{Ptuskin \& Zirakashvili}{2003}]{nll} Ptuskin V.~S., Zirakashvili V.~N., 2003, Astron. \& Astrophys., 403, 1
\bibitem[\protect\citeauthoryear{Sano et al.}{2010}]{sano} Sano H. et al., 2010, Astrophys.~J., 724, 1
\bibitem[\protect\citeauthoryear{Sano et al.}{2012}]{sano12} Sano T., Nishihara K., Matsuoka C., Inoue T., 2012, Astrophys.~J., 758, 2, 126
\bibitem[\protect\citeauthoryear{Sato et al.}{2018}]{sato18} Sato T., Katsuda S., Morii M., Bamba A., Hughes J.~P., Maeda Y., Ishida M., Fraschetti F., 2018, Astrophys.~J., 853, 46
\bibitem[\protect\citeauthoryear{Slane et al.}{1999}]{slane} Slane P., Gaensler B.~M., Dame T.~M., Hughes J.~P., Plucinsky P.~P., Green, A., 1999, Astrophys.~J., 525, 357
\bibitem[\protect\citeauthoryear{Skilling}{1971}]{skilling71} Skilling J., 1971, Astrophys.~J., 170, 265
\bibitem[\protect\citeauthoryear{Skilling}{1975}]{skilling} Skilling J., 1975, Mon.~Not.~R.~Astron.~Soc., 172, 557
\bibitem[\protect\citeauthoryear{Tanaka et al.}{2008}]{tanaka} Tanaka T. et al., 2008, Astrophys.~J., 685, 988
\bibitem[\protect\citeauthoryear{Truelove \& McKee}{1999}]{truelove} Truelove J.~K., McKee C.~F., 1999, Astrophys.~J.~Suppl., 120, 2
\bibitem[\protect\citeauthoryear{Uchiyama et al.}{2007}]{uchiyama} Uchiyama Y., Aharonian F.~A., Tanaka T., Takahashi T., Maeda Y., 2007, Nature, 449, 7162 
\bibitem[\protect\citeauthoryear{Uchiyama \& Aharonian}{2008}]{casa} Uchiyama Y., Aharonian F.~A., 2008, Astrophys.~J.~Lett., 667, L105
\bibitem[\protect\citeauthoryear{Villante \& Vissani}{2008}]{vissaniRxj} Villante F., Vissani F., 2008, Phys.~Rev.~D, 78, 103007
\bibitem[\protect\citeauthoryear{Wang, Qu \& Chen}{1997}]{china} Wang Z.~R., Qu Q.~Y., Chen ~Y., 1997, Astron. \& Astrophys., 318, L59
\bibitem[\protect\citeauthoryear{Zirakashvili \& Aharonian}{2010}]{felix} Zirakashvili V.~N., Aharonian F.~A., 2010, Astrophys.~J., 708, 2 
\bibitem[\protect\citeauthoryear{Zweibel \& Shull}{1982}]{in} Zweibel E.~G., Shull J.~M., 1982, Astrophys.~J., 259, 859
\end{thebibliography}



\bsp	
\label{lastpage}
\end{document}